\documentclass[a4paper,fleqn,usenatbib]{mnras}

\usepackage{newtxtext,newtxmath}
\usepackage[T1]{fontenc}
\usepackage{ae,aecompl}
\usepackage{color}

\usepackage{graphicx}	% Including figure files
\newcommand{\beq}{\begin{equation}}
\newcommand{\eeq}{\end{equation}}

\newcommand{\bfx}{\mbox{\boldmath{$x$}}}

\newcommand{\bfk}{\mbox{\boldmath{$k$}}}
\newcommand{\bfr}{\mbox{\boldmath{$r$}}}

\newcommand{\bftheta}{\mbox{\boldmath{$\theta$}}}
\newcommand{\bfell}{\mbox{\boldmath{$\ell$}}}

\newcommand{\kpc}{\mathrm{kpc}}
\newcommand{\Mpc}{\mathrm{Mpc}}
\newcommand{\Gpc}{\mathrm{Gpc}}

\newcommand{\rtrv}[1]{{\textcolor{black}{#1}}}

\title[Statistical modelling of the cosmological DM]{Statistical modelling of the cosmological dispersion measure} % with IllustrisTNG}

\author[R. Takahashi et al.]{
Ryuichi Takahashi,$^{1}$
Kunihito Ioka,$^{2}$
Asuka Mori$^{1}$
and Koki Funahashi$^{1}$
\\
$^{1}$Faculty of Science and Technology, Hirosaki University, 3 Bunkyo-cho, Hirosaki, Aomori 036-8561, Japan\\
$^{2}$Yukawa Institute for Theoretical Physics, Kyoto University, Kyoto 606-8502, Japan\\
}

\date{Accepted XXX. Received YYY; in original form ZZZ}

\pubyear{2020}

\begin{document}
\label{firstpage}
\pagerange{\pageref{firstpage}--\pageref{lastpage}}
\maketitle

%It should be a single paragraph not more than 250 words (200 words for Letters).\begin{abstract}
\begin{abstract}
We have investigated the basic statistics of the cosmological dispersion measure (DM)---such as its mean, variance, probability distribution, angular power spectrum and correlation function---using the state-of-the-art hydrodynamic simulations, IllustrisTNG300, for the fast radio burst (FRB) cosmology. 
To model the DM statistics, 
%As the DM statistics is fully determined by the cosmological free-electron distribution, 
we first measured the free-electron abundance and the power spectrum of its spatial fluctuations.
%in the state-of-the-art galaxy formation simulations, IllustrisTNG300.
The free-electron power spectrum turns out to be consistent with the dark matter power spectrum at large scales, but it is strongly damped at small scales ($\lesssim 1$Mpc) owing to the stellar and active galactic nucleus feedback.
%We then calibrate the theoretical model of DM statistics from the measured free-electron statistics.
The free-electron power spectrum is well modelled using a scale-dependent bias factor (the ratio of its fluctuation amplitude to that of the dark matter).
We provide analytical fitting functions for the free-electron abundance and its bias factor. % in the redshift range $z=0$--$5$.
We next constructed mock sky maps of the DM by performing standard ray-tracing simulations with the TNG300 data.
% based on a standard ray-tracing technique.
%the cosmological dispersion measure (DM) using the IllustrisTNG simulations. 
The DM statistics are calculated analytically from the fitting functions of the free-electron distribution, which agree well with the simulation results %such as the mean, variance and the angular power spectrum,
measured from the mock maps.
%The probability distribution of DM is highly positive skewed, consistent with the previous works. 
We have also obtained the probability distribution of source redshift for a given DM, which helps in identifying the host galaxies of FRBs from the measured DMs.
The angular two-point correlation function of the DM is described by a simple power law, $\xi(\theta) \approx 2400 (\theta/{\rm deg})^{-1} \, {\rm pc}^2 \, {\rm cm}^{-6}$, which we anticipate will be confirmed by future observations when thousands of FRBs are available. 
\end{abstract}


%%%%%%%%%%%%%%%%%%%%%%%%%%%%%%%%%%%%%%%%%%%%%%%%%%

%%%%%%%%%%%%%%%%% BODY OF PAPER %%%%%%%%%%%%%%%%%%

\section{Introduction}

A fast radio burst (FRB) is a radio pulse ($\sim$ ms wide) coming from a cosmological distance (see reviews by \cite{Cordes2019} and \cite{Petroff2019}).
After the first detection \citep{Lorimer2007}, more than hundreds of FRBs have been reported to date \citep{Petroff2016}.\footnote{FRB catalogue at \url{http://frbcat.org}}
Ongoing and future surveys such as ASKAP\footnote{\url{https://www.atnf.csiro.au/projects/askap/}}, CHIME\footnote{\url{https://chime-experiment.ca/}}, UTMOST\footnote{\url{https://astronomy.swin.edu.au/research/utmost/}}, FAST\footnote{\url{https://fast.bao.ac.cn/}}, STARE2 \citep{Boch2020} and SKA\footnote{\url{https://www.skatelescope.org/}} will detect thousands of events per year \citep[e.g.,][]{Connor2016,Hashimoto2020}.
%\rt{(need more discussion of experiments)}
Many FRB-progenitor models have been proposed, but the origins of FRBs are still obscure \citep[e.g.,][]{Popov2010,Totani2013,Kashi2013,Cordes2016,Murase2016,Metzger2017,Kumar2017,Levin2020,Lyubar2020,IokaZhang2020,Ioka2020}.\footnote{\url{https://frbtheorycat.org/index.php/Main_Page}}
More observations are needed to differentiate between them.
%Two kinds of FRBs (repeating or non-repeating) are known but these progenitors are still unknown.
From the frequency dependence of the arrival time from a FRB,  %with observed frequency $\nu$ is $\Delta t(\nu) \propto {\rm DM} \,\nu^{-2}$. 
the projected free-electron density along the line of light (i.e., the dispersion measure (DM)) can be measured.
Similarly, from the frequency dependence of the polarisation angle, the line-of-sight component of the magnetic field (i.e., the rotation measure (RM)) can also be measured. 
Because FRBs are extragalactic sources, these DMs and RMs directly map the cosmological free-electron distribution \citep{Ioka2003,Inoue2004} and the cosmic magnetic fields \citep[e.g.,][]{Akahori2016,Michilli2018}. 

The primordial abundance of baryons is currently measured to sub-percent-level accuracy by the cosmic microwave background (CMB) and Big Bang nucleosynthesis (BBN) \citep{Planck2018,Cooke2018}.
However, in the late-time universe, the baryon abundance and its spatial distribution are still poorly constrained by observations \citep[e.g.,][]{Fukugita2004,Shull2012}. 
%About $10 \, \%$ of baryons reside in virialized objects (such as galaxies, groups and clusters) in forms of stars and gas, while the rest are considered to be in the intergalactic medium (IGM) as warm or hot gas. 
About one-third of the baryons are still missing (the so-called `missing baryons'), although they are likely to be low-density ionised gas in the intergalactic medium (IGM). 
%It is quit important to detect these missing components. \rt{(<- need improvement)}
%An further observational constraint on the abundance at a low $z$ is desirable.
The cosmological DM is a powerful tool to probe for the missing baryons \citep{Ioka2003,Inoue2004}.
Very recently, \cite{MacQ2020} measured the baryon density from five host-galaxy-identified FRBs. 
Their result is independent of, but consistent with, the \textit{Planck} and BBN results. 
\cite{Keane2016} provided a similar constraint from a single event. 

%Therefore, the abundance of free electrons and its spatial clustering are important. 
Because FRBs and their DMs have unique cosmological properties, many cosmological applications have been proposed. 
The DMs of far-distant FRBs are a unique probe of cosmological reionisation \citep{Ioka2003,Inoue2004,Caleb2019,Dai2020}.
%Thanks to its compact size, 
Gravitational lensing of FRBs also enables searches for intervening compact objects that may constitute the dark matter \citep[e.g.,][]{Zheng2014,Munoz2016,Oguri2019,Jow2020,Liao2020}.
If the host galaxy is identified, the redshift--DM relation can constrain the dark energy models \cite[e.g.,][]{Gao2014,Zhou2014}.   
The angular auto-correlation of the DM directly maps free-electron clustering \citep[e.g.,][]{Masui2015,Shirasaki2017}, and its large-scale signal may contain primordial non-Gaussianity \citep{Reis2020}. 
The cross-correlation of the DM and foreground galaxies provides the free-electron distribution around the galaxies \citep{McQuinn2014,Shirasaki2017,Madh2019}, as well as helping to constrain the redshift distribution of the host galaxies \citep{RSM2019}. 
The cross-correlation of the DM and the thermal Sunyaev--Zel'dovich signal \citep[tSZ,][]{tSZ1970} gives further information about ionised gas, because the tSZ effect measures the projected electron pressure \citep{Munoz2018}.

Theoretical studies of the DM statistics have been based on the analytical halo model or hydrodynamic simulations because these are able to explore the non-linear free-electron distribution.
\cite{McQuinn2014} calculated the DM statistics (variance and probability distribution) from the halo model \citep[e.g.,][]{CS2002} for given model ingredients such as spatial halo clustering, halo mass function and free-electron density profile in halos. 
\cite{Madh2019} and \cite{Dai2020} computed the angular power spectrum of the DM based on the halo model.
Cosmological hydrodynamic simulations are the most reliable tools for investigating the free-electron distribution in the universe.
\cite{Dolag2015} studied the DM probability distribution based  on hydrodynamic simulations \citep[the Magneticum Pathfinder;][]{Dolag2016}.
\cite{Zhu2018} estimated the dispersion and scattering measures in the IGM using their cosmological hydrodynamic simulations.
\cite{Pol2019} made a full-sky map of the DM using the MICE ONION simulation \citep{Fosalba2008}, and they computed the mean, variance and probability distribution of the DM. 
That was a dark-matter-only simulation, and they assumed that the free electrons exactly trace the dark matter.  
\cite{Shirasaki2017} performed a similar analysis using their own dark-matter simulation. 
\cite{Jaros2019} recently studied the cosmological DM (its mean, variance and probability distribution) using a public hydrodynamic simulation, the original Illustris \citep{Vogel2014}.  
%They measured the spatial distribution of free electrons from the simulation.

The previous simulation studies did not compare their measurements with analytical predictions of the DM statistics (such as its variance and power spectrum), where the analytical solutions are useful for future data analyses.  
Previous analytical studies on the DM statistics assumed that free electrons exactly trace the underlying dark matter \citep{Masui2015,Shirasaki2017,RSM2019}, although this assumption breaks down at small scales ($\lesssim 1 \, \Mpc$), as shown in subsection 3.3.
The main purpose of the present work is to provide an analytical model for the DM statistics (such as its mean, variance, angular power spectrum and correlation function).
The analytical model is based on a standard two-point statistics.
Because the DM statistics are fully determined by the free-electron statistics, we first measure the the free-electron distribution from the latest cosmological hydrodynamic simulations, IllustrisTNG, the successor to Illustris \citep[e.g.,][]{Nelson2018a}.
We use the largest-box run from these TNG simulations (named TNG300, for which the side length of the cubic box is $L=205 \, h^{-1} \, \Mpc \, \simeq 300 \, \Mpc)$, which is suitable for cosmological studies. 
We measure the free-electron abundance and the power spectrum of its spatial fluctuations over a wide range of redshifts ($z=0$--$5$) and scales ($\approx 0.1$--$200 \, h^{-1} \, \Mpc$) in TNG300.
We then make fitting functions for them to model the free-electron distribution.
The DM statistics are calculated analytically using these fitting functions.
%Our model is based on standard two-point statistics, in which the free-electron distribution (i.e., its abundance and power spectrum) is calibrated from TNG300 using simple fitting functions. 
%Because only hydrodynamic simulations are able to explore its non-linear distribution.
%First, we model the free-electron distribution because 
%Our model for the DM statistics includes these fitting functions.
%In our theoretical model, the free-electron statistics is modelled using these fitting functions.
We next construct mock sky maps of the DM using the TNG300 data and measure the DM statistics from them to check the accuracy of the analytical model.
% compute the DM statistics.
%and easily apply it for another cosmological model and free-electron distribution model. 
The three spatial-resolution runs in TNG300 are used to check the numerical convergence of the results. 
The presented model is applicable, in principle, for other cross-correlations, such as DM--galaxy, DM--weak lensing and DM--tSZ cross-correlations. 
As thousands of FRBs will be available in the relatively near future, 
we expect this kind of statistical study to be required.
%we expect this kind of statistical study to be very useful.} 
%We use public data of the IllustrisTNG simulations, %The simulation data is publicly available \citep{Nelson2018}.
%Although the TNG team performed three different box-size simulations, 
Throughout this paper, we mainly study the cosmological DM (i.e., excluding contributions from the Milky Way and host galaxies). 
 
The rest of this paper is organised as follows:
Section 2 introduces the theory of two-point DM statistics.
Section 3 measures the free-electron abundance and its power spectrum in the TNG300 data and provides fitting functions for them.
Section 4 describes a procedure for making mock sky maps of the DM.
Section 5 presents our main results: comparisons between the simulation results measured from the mock maps and analytical predictions.
Section 6 discusses the host-galaxy contribution and provides comparisons with other hydrodynamic simulations. 
Finally, section 7 summarises this work.  

Throughout this paper, we adopt a cosmological model consistent with the \textit{Planck} 2015 best-fit flat $\Lambda$CDM model \citep{Planck2015}: matter density $\Omega_{\rm m}=1-\Omega_{\Lambda}=0.3089$, baryon density $\Omega_{\rm b}=0.0486$, Hubble parameter $h=0.6774$, spectral index $n_{\rm s}=0.9667$, and amplitude of the matter density fluctuations on the scale of $8 \, h^{-1} \, \Mpc$ $\sigma_8=0.8159$.
This model is the same as that adopted in the TNG simulations.
All physical quantities (such as length, wavenumber and number density) will be given in comoving units.  %unless otherwise stated.

\section{Theory of the cosmological dispersion measure}

This section presents the theoretical basics of the cosmological DM: the mean and fluctuations (subsection 2.1) and the two-point statistics (subsection 2.2). 

\subsection{The mean and fluctuations}

\begin{figure}
 \includegraphics[width=1.1 \columnwidth]{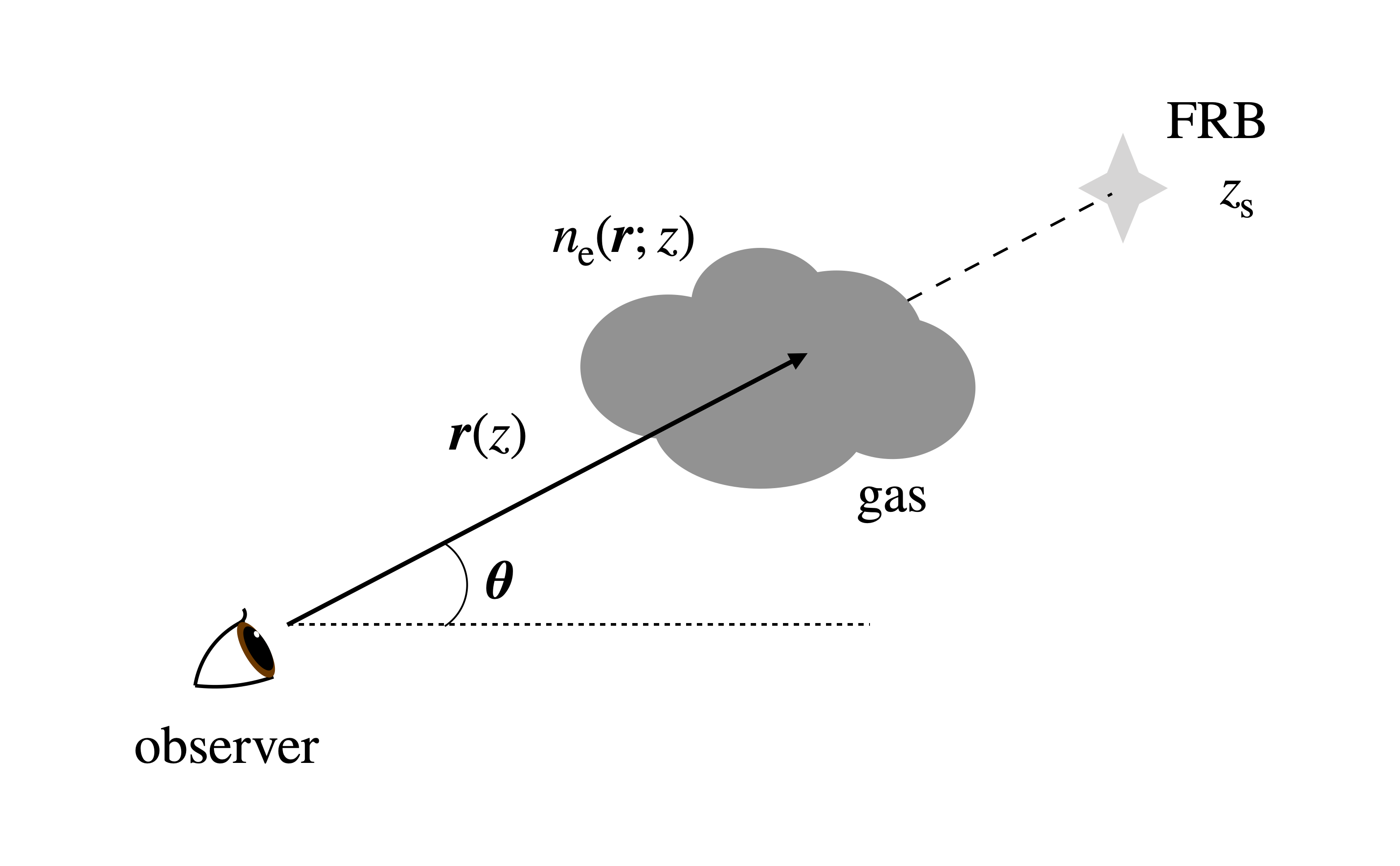}
 \vspace{-0.5cm}
 \caption{Schematic configuration of the observer, ionised gas and an FRB: $\bftheta$ is the angular position of the FRB, $\bfr$ is a vector along the line of sight and $n_{\rm e}$ is the number density of free electrons at $\bfr$.}
\label{fig_FRB_config}
\end{figure}

Three major components contribute to the observed DM: the Milky Way, the host galaxy and the intervening cosmological medium.
The Milky Way contribution can be inferred from the Galactic free-electron distribution, which is modelled by pulsar measurements \citep[e.g., the NE2001 model:][]{Cordes2002}.
The host-galaxy contribution decreases for more distant sources in proportion to $(1+z_{\rm s})^{-1}$, where $z_{\rm s}$ is the source redshift, due to cosmological time dilation and the Doppler frequency shift \citep[if its intrinsic property in the rest frame does not evolve with time, e.g.,][]{Zhou2014}.
In contrast, the cosmological contribution increases roughly in proportion to $z_{\rm s}$ \citep[e.g.,][]{Ioka2003}, and it exceeds the host-galaxy contribution for $z_{\rm s} \gtrsim 0.3$.
Therefore, throughout this paper, we mainly consider the cosmological contribution, and hereafter, DM refers to that alone. % unless otherwise stated.
The host-galaxy contribution will be briefly discussed in subsection 6.1.

%, i.e., ${\rm DM}_{\rm total}={\rm DM}_{\rm MW}+{\rm DM}_{\rm host}+{\rm DM}_{\rm cosmo}$.
We consider an FRB at an angular position $\bftheta = (\theta_1,\theta_2)$ on the sky and redshift $z_{\rm s}$, as shown in Fig. \ref{fig_FRB_config}.
%Here, $\bftheta = (\theta_1,\theta_2)$ is defined as an angular vector from a fixed direction (as denoted by the dashed line in the figure).
The vector $\bfr$ points to the intervening gas at $z$, and its absolute value is the comoving distance
\beq
 r(z)=\int_0^{z} \!\! \frac{c dz^\prime}{H(z^\prime)},
 \label{r-z_relation}
\eeq
where $H(z)$ is the Hubble expansion rate. 
%The intervening gas cloud contains free electrons.
Denoting the number density of free electrons at $\bfr$ and $z$ by $n_{\rm e}(\bfr;z)$,
the DM is obtained by integrating $n_{\rm e}$ along the line of sight  \citep[e.g.,][]{Ioka2003,Inoue2004}:
\beq
  {\rm DM}(\bftheta;z_{\rm s}) = \int_{0}^{z_{\rm s}} \!\!\! \frac{c dz}{H(z)}  n_{\rm e} (\bfr;z) (1+z).
\label{dm}
\eeq
Note that the number density $n_{\rm e}$ is given in comoving units. 
%We assume $|\bftheta| \ll 1$.

The number density $n_{\rm e}$ can be decomposed into its spatial mean $\bar{n}_{\rm e}$ and fluctuations $\delta_{\rm e}$:
\beq
 n_{\rm e} (\bfr;z) = \bar{n}_{\rm e} (z) \left[  1 + \delta_{\rm e} (\bfr;z) \right].
 \label{ne}
\eeq
The spatial average of the second term vanishes: $\langle \delta_{\rm e} \rangle =0$.
As for the first term, the total number density of electrons (including both free electrons and those bound to atoms) in the universe is 
\beq
  \bar{n}_{\rm e,total} = \left( X_{\rm p}+ \frac{1}{2} Y_{\rm p} \right) \frac{\bar{\rho}_{\rm b}}{m_{\rm p}},
\eeq
where $\bar{\rho}_{\rm b}$ is the comoving cosmological baryon density, and $m_{\rm p}$ is the proton mass \citep[e.g.,][]{DZ2014}. 
The quantities $X_{\rm p}$ and $Y_{\rm p}$ denote the primordial mass fractions of hydrogen and helium, respectively, and are set to $X_{\rm p}=1-Y_{\rm p}=0.76$ to be consistent with TNG.
We ignore the time evolution of $\bar{n}_{\rm e,total}$ due to stellar nucleosynthesis, because it is negligibly small.
Introducing the free-electron fraction at $z$, $f_{\rm e}(z)$, the free-electron number density is written as
\beq
  \bar{n}_{\rm e}(z) = f_{\rm e}(z) \, \bar{n}_{\rm e,total},
\label{ne_mean}
\eeq
where $f_{\rm e}=1$ corresponds to full ionisation.
%In their first studies, \cite{Ioka2003} and \cite{Inoue2004} simply assumed $f_{\rm e}=1$.
After hydrogen and helium were fully ionised at $z \sim 3$, $f_{\rm e}$ is assumed to be close to unity.  
However, current observational constraints on $f_{\rm e}$ still have a large variation \cite[$f_{\rm e} \simeq 0.7$--$1$, e.g.,][]{Fukugita2004,Shull2012,McQuinn2016,Walters2019,Li2020}.
Note that $f_{\rm e}$ in Eq. (\ref{ne_mean}) includes all free electrons, both inside and outside of intervening galaxies.
In other words, $f_{\rm e}$ is the spatial mean fraction averaged over all galaxies and the IGM.
In this paper, we do not introduce the free-electron fraction in IGM, $f_{\rm IGM}$.
One reason is that $f_{\rm IGM}$ depends on the boundary between the galaxies and the IGM, and that boundary is ambiguous.
Another reason is that some FRB signals may pass through an intervening galaxy; this probability may be low, but it gives a large DM.
We will measure $f_{\rm e}$ from the TNG300 simulations in section 3.

Similarly to $n_{\rm e}$, the DM can be decomposed into two terms,
\beq
 {\rm DM} (\bftheta; z_{\rm s}) = \overline{\rm DM} (z_{\rm s}) + \delta {\rm DM} (\bftheta; z_{\rm s}).
\eeq  
The mean and fluctuations of the DM can be written from Eqs. (\ref{dm})--(\ref{ne_mean}) in the forms
\begin{align}
 \overline{\rm DM}(z_{\rm s}) &=  \int_0^{z_{\rm s}} \!\! \frac{c dz}{H(z)} W(z),  
 \label{mean_DM}  \\
 \delta {\rm DM}(\bftheta; z_{\rm s}) &=  \int_0^{z_{\rm s}} \!\! \frac{c dz}{H(z)} W(z) \delta_{\rm e}(\bfr;z), 
 \label{DM_fluct}
\end{align}
with a kernel
\beq
W(z) = \frac{\bar{\rho}_{\rm b}}{m_{\rm p}} \left( X_{\rm p} + \frac{1}{2} Y_{\rm p} \right) f_{\rm e}(z) (1+z).
%W(z) = \frac{3 H_0^2}{8 \pi} \frac{\Omega_{\rm b}}{m_{\rm p}} \left( X + \frac{1}{2} Y \right) f_{\rm e}(z) (1+z).
\label{w_kernel}
\eeq
%The angular average of $\delta {\rm DM}$ is zero, $\langle \delta {\rm DM} \rangle=0$.
The mean baryon density is rewritten as $\bar{\rho}_{\rm b}=\Omega_{\rm b} \, \rho_{\rm cr} = 3 H_0^2 \Omega_{\rm b}/(8 \pi G)$, where $\rho_{\rm cr}$ is the cosmological critical density.

\subsection{The two-point statistics}

%decomposed into a parallel and perpendicular components, $\bfr=(r \bftheta,r)$.
This subsection discusses the angular correlation function and its Fourier transform (i.e., the power spectrum) of the DM fluctuations. 
Previously, several authors have studied the angular power spectrum of the DM \citep[e.g.,][]{Masui2015,Shirasaki2017,Madh2019,Dai2020}.
%, and \citet{Masui2015} studied the angular clustering of FRB host galaxies. 
Here, we simply summarise their results.\footnote{A detailed discussion of the two-point statistics of projected random fields is found in, e.g., section 2.4 of \citet{BS2001} and section 9.1 of \citet{Dodelson2003}.}
%,Schneider2006}.
%studied in weak lensing. 

The angular correlation function of the DM between $\bftheta_1$ and $\bftheta_2$ at the same source redshift $z_{\rm s}$ is defined as
\beq
  \xi(\theta_{12};z_{\rm s}) \equiv \langle \delta {\rm DM}(\bftheta_1;z_{\rm s})  \, \delta {\rm DM}(\bftheta_2;z_{\rm s}) \rangle.
  \label{xi}
\eeq
%where $\theta_{12}=|\bftheta_1-\bftheta_2|$. 
Because of the isotropy of the universe, the correlation function is a function of the separation $\theta_{12}=|\bftheta_1-\bftheta_2|$.
Throughout this paper, we assume $|\theta_{12}| \ll 1$, i.e. the flat-sky approximation is valid.
From Eqs. (\ref{DM_fluct}) and (\ref{xi}), under the Limber and the flat-sky approximations, the correlation function reduces to
\beq
  \xi(\theta_{12};z_{\rm s}) = \frac{1}{2 \pi} \int_0^{z_{\rm s}} \!\! \frac{c dz}{H(z)} W^2(z) \int_0^{\infty} \!\! dk k P_{\rm e}(k;z) J_0 \left( \theta_{12} k r(z) \right),
  \label{xi2}
\eeq
where $J_0$ is the zero-th order Bessel function, and $k$ is the wavenumber of the density fluctuations.
The power spectrum of the free-electron fluctuations is defined as
\beq
  P_{\rm e}(k;z) \left( 2 \pi \right)^3 \delta_{\rm D}^3 (\bfk+\bfk^\prime) \equiv \langle \tilde{\delta}_{\rm e}(\bfk;z) \tilde{\delta}_{\rm e}(\bfk^\prime;z) \rangle,
\eeq
where $\tilde{\delta}_{\rm e}(\bfk;z)$ is the Fourier transform of $\delta_{\rm e}(\bfr;z)$, and $\delta_{\rm D}$ is the Dirac delta function.

The Fourier transform of the DM fluctuations is given by
\beq
  \widetilde{\delta {\rm DM}} (\bfell ;z_{\rm s}) = \int \! d^2 \bftheta \, \delta {\rm DM} (\bftheta ;z_{\rm s})  \, {\rm e}^{-{\rm i} \, \bfell \! \cdot \bftheta},
\label{DM_fft}
\eeq
where $\bfell=(\ell_1,\ell_2)$ is the two-dimensional vector of multipole moments. 
Similarly to $P_{\rm e}(k;z)$, the angular power spectrum of the DM is defined as
\beq
   C_\ell (z_{\rm s}) \left( 2 \pi \right)^2 \delta_{\rm D}^2 (\bfell+\bfell^\prime) \equiv \langle \widetilde{\delta {\rm DM}} (\bfell ;z_{\rm s}) \widetilde{\delta {\rm DM}} (\bfell^\prime ;z_{\rm s})  \rangle.
\label{cl_def}
\eeq
From the above equations (\ref{xi})--(\ref{cl_def}), the angular power spectrum is obtained as
\begin{align}
  C_\ell (z_{\rm s}) &= \int \! d^2 \bftheta \, \xi(\theta;z_{\rm s}) \, {\rm e}^{-{\rm i} \, \bfell \! \cdot \bftheta},\nonumber  \\
   &= \int_0^{z_{\rm s}} \!\! \frac{c dz}{H(z)} \frac{W^2(z)}{r^2(z)} P_{\rm e} \left(k=\frac{\ell}{r(z)};z \right).  
  \label{cl_DM}
\end{align}
This equation relates the 3D power spectrum of the free electrons to the 2D power spectrum of the DM. 

The variance of the $\rm DM$ is simply obtained by setting $\bftheta_{1}=\bftheta_2$ in Eqs. (\ref{xi}) and (\ref{xi2}):
\begin{align}
  \sigma_{\rm DM}^2 (z_{\rm s}) &\equiv \langle \left[ \delta {\rm DM} (\bftheta;z_{\rm s}) \right]^2 \rangle, \nonumber \\
  &= \frac{1}{2 \pi} \int_0^{z_{\rm s}} \!\! \frac{c dz}{H(z)} W^2(z) \int_0^{\infty} \!\! dk k P_{\rm e}(k;z).
  \label{dm_var}
\end{align}
This is consistent with the analytical result in \citet[][their section 2]{McQuinn2014}.
Theoretical models of the ionised fraction $f_{\rm e}(z)$ and the power spectrum $P_{\rm e}(k;z)$ are required to compute the above two-point statistics.
We will calibrate these functions using TNG300 in the next section.

\section{Calibration with TNG300}

This section briefly introduces the TNG simulations (subsection 3.1) and then measures the free-electron fraction $f_{\rm e}(z)$ (subsection 3.2) and the power spectrum $P_{\rm e}(k;z)$ (subsection 3.3). 
%from the TNG dataset.

\subsection{The TNG simulations}

\begin{table*}
 \centering
 \caption{Summary of the TNG300 simulations used in this paper: the numbers of baryon and dark-matter particles ($N_{\rm baryon},N_{\rm dark}$), the average masses of baryon and dark-matter particles ($m_{\rm baryon}$, $m_{\rm dark}$), the minimum gravitational softening length of the gas cells ($\epsilon_{\rm gas,min} $),  and the mean size of the gas cells ($r_{\rm gas} \equiv L/N_{\rm baryon}^{1/3}$). The upper three runs follow both the gravitational evolution and astrophysical processes, while the bottom one follows only the former. The side length of the simulation box is $L=205 \, h^{-1} \, \Mpc$ in all runs. }
 \label{table_TNG}
 \begin{tabular}{lcccccc}
  \hline 
   & $N_{\rm baryon}$ &  $N_{\rm dark}$ & $m_{\rm baryon} (h^{-1} {\rm M}_\odot)$ & $m_{\rm dark} (h^{-1} {\rm M}_\odot)$ &  $\epsilon_{\rm gas,min} (h^{-1} \kpc)$ & $r_{\rm gas}$ $\! (h^{-1} \, \kpc)$  \\
  \hline
   TNG300-1 & $2500^3$ &  $2500^3$ & $7.4 \times 10^6$  & $4.0 \times 10^7$ & $0.25$ & $82$ \\
   TNG300-2 & $1250^3$ &   $1250^3$ & $6.0 \times 10^7$ & $3.2 \times 10^8$ & $0.5$ &  $164$ \\
   TNG300-3 & $625^3$ &  $625^3$ &  $4.8 \times 10^8$ & $2.5 \times 10^9$  & $1.0$ &   $328$ \\ \hline
   TNG300-1-Dark & -- &  $2500^3$ &  --   &  $4.7 \times 10^7$ &  -- & -- \\  \hline
 \end{tabular}
\end{table*}

\begin{table}
 \centering
 \caption{Output redshift $z$, comoving distance $r(z)$ and snapshot number in the TNG dataset.}
 \label{table_list}
 \begin{tabular}{ccc}
  \hline
  $z$ & $r(z) \, (h^{-1} {\rm Mpc})$ & snapshot  \\
  \hline
   0 & 0 & 99 \\
   0.1 & 293 & 91 \\
   0.2 & 571 & 84 \\
   0.3 &  834 & 78 \\
   0.4 & 1083 & 72 \\
   0.5 & 1318 & 67 \\
   0.7 & 1747 & 59 \\
   1 & 2301 & 50 \\
  1.5 & 3034 & 40 \\
   2 &  3599 & 33 \\
   3 & 4411 & 25 \\
   4  & 4973 & 21 \\
   5  & 5390 & 17 \\
   6  &  5716 & 13  \\
   7 &  5978  & 11 \\
   8 &  6196  & 8 \\
  \hline
 \end{tabular}
\end{table}

\begin{table}
 \centering
 \caption{Mass fractions of gas, stars and super-massive black holes to the total baryons measured in TNG300-1. The values are given in percentages (i.e., $f_{\rm \, gas}+f_{\rm \, star}+f_{\rm \, bh}=100 \, \%$). 
 The gas is further decomposed into neutral and ionised hydrogen (${\rm H}_0$ and ${\rm H}^+$) and helium (He), which satisfy $f_{\rm gas} \simeq f_{\, {\rm H}_0}+f_{\, {\rm H}^+}+f_{\, {\rm He}}$.}
 \label{table_baryon}
 \begin{tabular}{ccccccc}
  \hline
  $z$ & $f_{\rm \, gas}$ & $f_{\, {\rm H}_0}$ & $f_{\, {\rm H}^+}$ & $f_{\rm \, He}$ & $f_{\rm \, star}$ & $f_{\rm \, bh} $    \\
  \hline
   0 & 96.8 & 1.4 & 71.9 &	23.3 & 3.2 & 0.02 \\
   1 & 97.7 & 0.9 & 73.2 &	23.5 & 2.3 & 0.01 \\
   %2 & 98.6 & 0.9 & 74.0 & 23.7 & 1.4 & $9 \times 10^{-3}$ \\
   3 & 99.3 &	0.9 & 74.6 & 23.8 & 0.7 & $5 \times 10^{-3}$ \\
   %4 & 99.6 & 1.1 & 74.6 & 23.9 & 0.4 & $2 \times 10^{-3}$ \\
   5 & 99.8 &	1.3 & 74.5 &	24.0 & 0.2 &	$1 \times 10^{-3}$ \\
   6 & 99.9 &	74.0 & 1.9 &	24.0 & 0.1 & $<10^{-3}$ \\
   7 & 100.0 & 75.0 &	1.0 & 24.0 & 	0.06 & $<10^{-3}$ \\
   8 & 100.0 & 75.5 &	0.4 & 24.0 & 0.03 & $<10^{-3}$ \\
  \hline
 \end{tabular}
\end{table}

We investigate the spatial distribution of free electrons in the universe using the TNG dataset\footnote{The simulation data is available at \url{http://www.tng-project.org}} \citep{Marinacci2018,Naiman2018,Nelson2018a,Pillep2018,Springel2018}.
The simulations follow the gravitational clustering of matter (dark matter and baryons) as well as astrophysical processes such as star and galaxy formation, gas cooling, and stellar and active galactic nucleus (AGN) feedback.   
The gravitational evolution and magneto-hydrodynamic processes were computed with the moving-mesh code \texttt{AREPO} \citep{Springel2010}.
The simulations incorporate astrophysical processes in a subgrid model, thereby enabling them to follow the processes of galaxy formation and evolution. 
The TNG project produced three sets of simulations in different-sized cubic boxes, with three mass resolutions for each box size.
Here, we used the largest box (referred to as TNG300), with side length $L=205 \, h^{-1} \, \Mpc \, (\simeq 300 \, \Mpc)$, because our interest is the large-scale distribution of free electrons.
To check the numerical convergence, we used the three resolutions from high to low (referred to as TNG300-1 to -3, respectively).
This box contains the same number of dark-matter and baryon particles.
The number of particles and the mass resolution are listed in Table \ref{table_TNG}.
The TNG team also performed dark-matter-only runs, in which the number of dark-matter particles was the same as in TNG300. 
In this case, the N-body particles represent both components (baryons and dark matter), but the simulations follow the gravitational evolution only.
Such simulations help to see the impact of dark matter on the free-electron clustering.
Here we used the highest-resolution run (named TNG300-1-Dark).
%The cosmological model is based on the \textit{Planck} 2015 flat $\Lambda$CDM \citep{Planck2015}.
The TNG team have released the simulation data at $20$ redshifts in the range $z=0$--$12$ (named ``full'' snapshots).
In this paper, we used all the datasets up to $z=8$, as listed in Table \ref{table_list}.
The first column is the redshift $z$, the second is the comoving distance to $z$ and the third refers to the TNG snapshot number. % from $8$ to $99$ (corresponding to $z=8$ to $0$).

Each baryon particle has one of three forms: gas, star or super-massive black hole.
Free electrons are contained only in the gas particles.
%Most of the baryons are gas.
At the initial redshift ($z=127$), all the baryon particles are gas. 
As time evolves, the gas falls into the halos, and star formation begins in high-density regions \citep{Pill2018b}.
Some gas particles then convert to stars or black holes.
However, even at $z=0$, most of the baryon particles are still gas (the gas mass fraction is $>96 \%$).  
The time evolution of each mass fraction measured in TNG300-1 is summarised in Table \ref{table_baryon}.
The mass fraction is obtained from the total mass of each component in the box divided by the total baryonic mass ($=\bar{\rho}_{\rm b} L^3$).
The TNG team followed the time evolution of the atomic abundances of H, He and seven other species (C, N, O, Ne, Mg, Si and Fe).
The public data contains the atomic abundance in each gas particle.
For hydrogen, the data includes neutral (${\rm H}_0$) and ionised (${\rm H}^+$) fractions. 
In Table \ref{table_baryon}, the gas-mass fraction ($f_{\rm \, gas}$) is further decomposed into neutral and ionised hydrogen ($f_{\,{\rm H}_0}$ and $f_{\,{\rm H}^+}$) and helium ($f_{\rm \, He}$), where $f_{\rm \, He}$ includes both neutral and ionised states.
The mass fraction of elements heavier than He is negligible.
The hydrogen is ionised abruptly at the epoch of cosmic reionisation (between $z=5$ and $6$).
%In this paragraph, we compare the TNG mass fraction in Table \ref{table_baryon} with observations.
We comment that the mass fraction of stars ($f_{\rm star}$) reaches $3.2 \, \%$ at $z=0$, which is slightly smaller than the observed values $f_{\rm star} = 6.0 \pm 1.3 \, \%$ \citep{Fukugita2004} and $7 \pm 2 \, \%$ \citep{Shull2012,Nicastro2018}.
Therefore the gas fraction $f_{\rm gas}$ and the resulting free-electron fraction $f_{\rm e}$ in TNG300-1 may be overestimated by approximately a few percent.
%The mass fraction of ${\rm H}_0$ is $1.4 \, \%$ at $z=0$, which is consistent with observed total abundance of %HI gas and hydrogen molecule of $1.3 \pm 0.2 \, \%$ \citep{Fukugita2004}.

%The IllustrisTNG team also performed dark-matter-only (dmo) runs.
%By comparing the simulations in the presence and absence of baryons, we can single out the impact of baryons on matter clustering. 

Previous work \citep{Jaros2019} on the DM used the lowest-resolution run of Illustris: the side length of the box is $75 \, h^{-1} \, \Mpc$, and it contains $2 \times 455^3$ baryon and dark-matter particles.
Therefore, TNG300 has better mass resolution and a larger simulation volume. 
Their work did not check the numerical convergence among the different resolutions.
Illustris is known to predict AGN feedback that is too strong \citep[e.g.,][]{Chisari2019}, which may affect the free-electron distribution in the halos.

\subsection{Free-electron abundance}

\begin{figure}
 \includegraphics[width=1.15 \columnwidth]{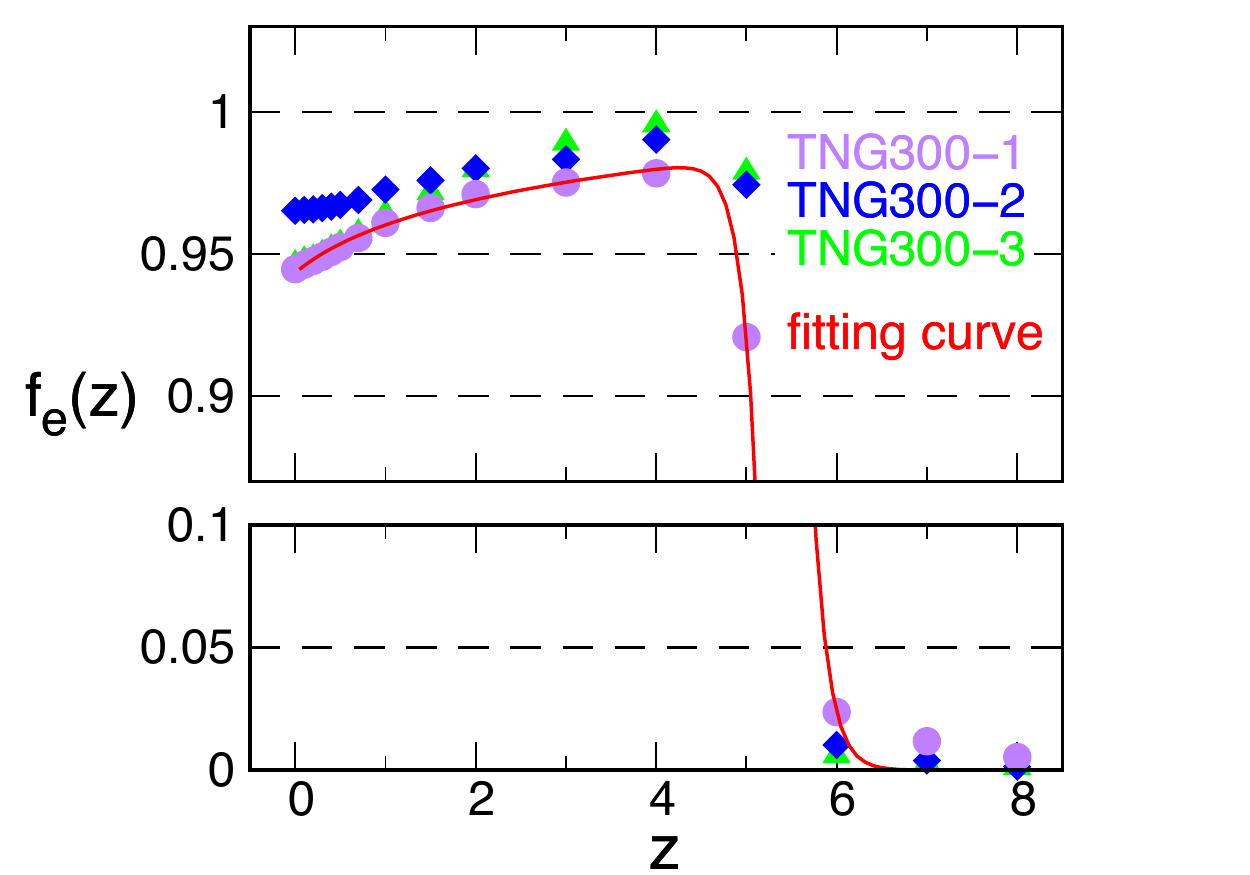}
 \caption{Time evolution of the free-electron fraction measured in TNG300. The purple, blue and green symbols represent the results from TNG300-1, -2 and -3 (from high- to low-resolution runs), respectively. The red curve is the fit to TNG300-1 given in Eq. (\ref{fe_fit}).}
\label{fig_mean_ne}
\end{figure}

The TNG team also provided the abundance of free electrons in each gas particle, which is the total free-electron abundance for all atoms (not only for hydrogen). 
%Therefore, we follow the spatial distribution of free electrons.
By summing up all the gas particles in each snapshot, we obtained the number density of free electrons and its fraction, $\bar{n}_{\rm e}(z)$ and $f_{\rm e}(z)$, as defined in Eq. (\ref{ne_mean}). 
Here we measured $f_{\rm e}(z)$ at the $16$ redshifts in the range $z=0$--$8$ listed in Table \ref{table_list}.
The result is plotted in Fig. \ref{fig_mean_ne}. % plots $f_{\rm e}(z)$ measured in TNG300.
At high $z \, (\gtrsim 6)$, the gas is still neutral (i.e., $f_{\rm e} \simeq 0$).
The fraction $f_{\rm e}$ rises abruptly at the epoch of hydrogen reionisation ($z \sim 6$) and increases further at the epoch of helium reionisation ($z \sim 4$).
At relatively low $z \, (\lesssim 3)$, $f_{\rm e}$ decreases slightly because some fraction of the electrons becomes confined in stars and black holes (see Table \ref{table_baryon}). 
A small fraction of the electrons is in neutral hydrogen (HI and ${\rm H}_2$) in galaxies \cite[$f_{{\rm H}_0} \sim 1 \, \%$ in Table \ref{table_baryon}; the cosmological HI distribution was recently studied using hydrodynamic simulations in, e.g.,][]{Vill2018,Ando2019}.  

The TNG300-1 result can be fitted by
\beq
  f_{\rm e}(z)=a \left( z+b \right)^{0.02} \left[ 1-\tanh \left\{ c (z-z_0) \right\} \right],
  \label{fe_fit}
\eeq
with $a=0.475, b=0.703, c=3.19$ and $z_0=5.42$. 
Here, $z_0$ corresponds to the epoch of hydrogen reionisation.
For $z \gg z_0$, $f_{\rm e}$ approaches zero. 
On the other hand, for $z \to 0$, $f_{\rm e} \rightarrow 2a \simeq 0.95$.
This fit agrees with the TNG300-1 results to within a deviation $\Delta f_{\rm e}=0.012$ in the range $z=0$--$8$.
There are few-percent deviations among the different-resolution runs, and thus, this fit has the same level of error.
Note that our $f_{\rm e}$ corresponds to the quantity $f_{\rm ion}$ defined in a previous work 
%using the original Illustris simulation with the lowest resolution run 
\citep{Jaros2019}.
Their result is slightly higher than ours, but the difference is very small ($f_{\rm ion} \simeq 0.98$--$0.99$ in the range $z=0$--$4$, see their Table 2).   

We comment that the TNG simulations followed the ionising state of IGM using the time-dependent spatially-uniform UV background radiation (instead of solving radiative transfer equations) with corrections for self shielding in dense gas.
This ionising background started at $z=6$, and thus the results of $f_{\rm e}$ at $z \geq 6$ should be considered with caution (see section 5.2 of \citet{Nelson2018}; section 2.1.2 of \citet{Pill2018b}).
In the following, the simulation data up to $z=5$ will be used.  %, therefore this does not affect our results.   

\subsection{Free-electron power spectrum}

\begin{figure*}
 \includegraphics[width=2 \columnwidth]{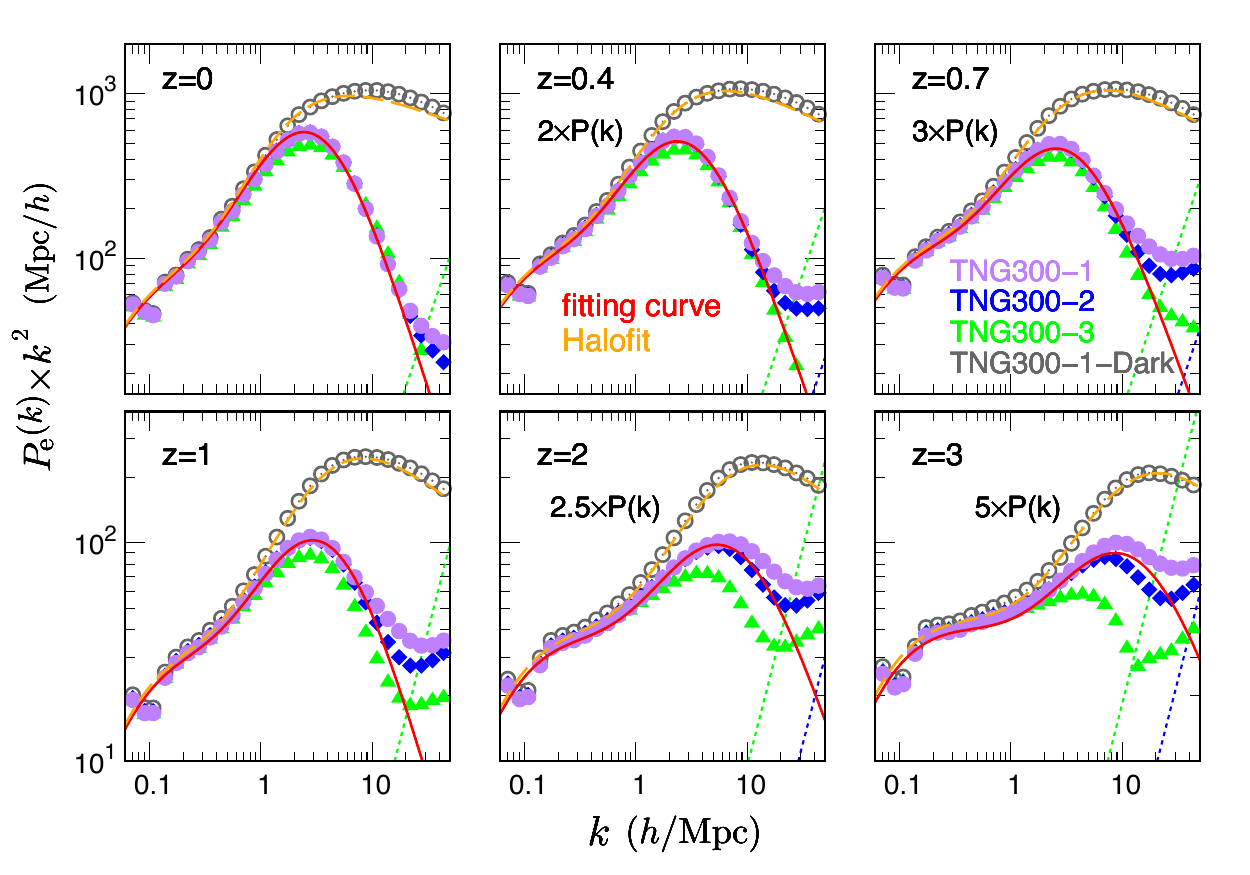}
 \vspace{0cm}
 \caption{Free-electron power spectra $P_{\rm e}(k)$ at $z=0$--$3$ measured from TNG300. The purple, blue and green symbols are the results from TNG300-1, -2 and -3, respectively. The grey circles are the matter power spectrum $P_{\rm dmo}(k)$ in the dark-matter-only run (TNG300-1-Dark). The dashed-orange curves are Halofit results for non-linear $P_{\rm dmo}(k)$ \citep{Takahasi2012}. The solid red curves are our fits: the free-electron bias factor, $b_{\rm e}^2(k)$ in Eq. (\ref{bias_fit}), times the Halofit. The green- and blue-dotted lines are the shot noise for TNG300-3 and -2, respectively. In the middle and right panels, the results are multiplied by factors of $2$--$5$ (as indicated in each panel) to make the presentation clearer.}
\label{fig_pk}
\end{figure*}

\begin{figure*}
 \vspace{-2cm}
 \includegraphics[width=2 \columnwidth]{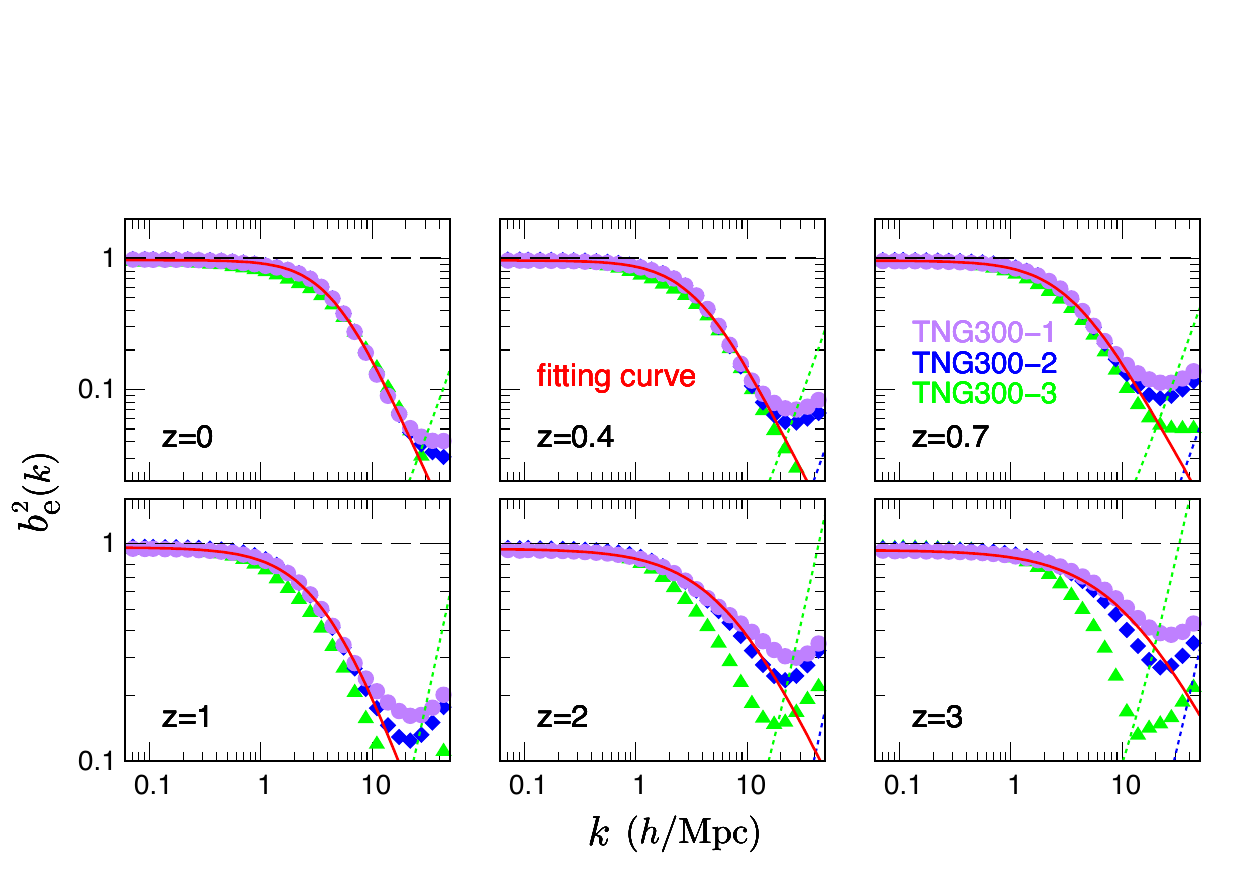}
 \vspace{0cm}
 \caption{Similar to Fig. \ref{fig_pk}, but showing the free-electron bias factor defined as $b_{\rm e}^2(k)=P_{\rm e}(k)/P_{\rm dmo}(k)$. The red curves are the fits to TNG300-1 given by Eq. (\ref{bias_fit}). The dotted lines denote the shot noise, and the horizontal dashed lines are $b_{\rm e}=1$.}
\label{fig_bias}
\end{figure*}

We next measured the power spectrum of free electrons in TNG300 following the standard procedure \citep[see, e.g.,][their section 2.2]{Springel2018}.
The TNG team provided the mass of free electrons in each gas particle. 
%We obtained the positions and free-electron masses for all gas particles from the simulation data.
To measure the density contrast, we assigned the free-electron mass to $1024^3$ regular grid cells in the box using the cloud-in-cell (CIC) interpolation with the interlacing scheme \citep[e.g.,][]{Jing2005,Sefusatti2016}.
The Fourier transform of the density field $\tilde{\delta}_{\rm e}(\bfk)$ was then obtained with fast Fourier transform (FFT)\footnote{\textsc{FFTW} (Fast Fourier Transform in the West) at \url{http://www.fftw.org}.}.
To explore smaller scales, we also employed the folding method \citep{Jenkins1998}, which folds the particle positions $\bfx$ into a smaller box of side length $L/10$ by replacing $\bfx$ with $\bfx \% (L/10)$ (where $a \% b$ denotes the reminder of $a/b$).
This procedure effectively increases the spatial resolution by $10$ times.
%The wavenumber has discrete number, $\bfk=(2 \pi/L) \bfn$ where the components of $\bfn$ are integers. 
The minimum and maximum wavenumbers in the $1024^3$ cells are $k_{\rm min}= 2 \pi/L = 0.025 \, h \, \Mpc^{-1}$ (where $L=205 \, h^{-1} \, \Mpc$) and $k_{\rm max}=512 \, k_{\rm min} = 12.9 \, h \, \Mpc^{-1}$, respectively. The folding scheme enlarges $k_{\rm max}$ by $10$ times. 
The power spectrum is reliable up to the particle Nyquist wavenumber, which is determined by the mean separation of the gas particles $r_{\rm gas}$ in Table \ref{table_TNG}: $k_{\rm Nyq}=\pi/r_{\rm gas}$. 
The values of $k_{\rm Nyq}$ are $38.3 \, (19.1 ~{\rm and}~ 9.6) \, h \, \Mpc^{-1}$ for TNG300-1 (-2 and -3). 

The power spectrum is measured as
\beq
 P_{\rm e}(k) = \frac{1}{N_{\rm mode}} \sum_{|\bfk^\prime| \in k} \left| \, \tilde{\delta}_{\rm e}(\bfk^\prime) \right|^2,
 \label{PS}
\eeq
where the summation is performed in the spherical shell $k-\Delta k/2<|\bfk^\prime|<k+\Delta k/2$ and $N_{\rm mode}$ is the number of Fourier modes in the shell with bin-width ($\Delta \log_{10} k=0.1$). 
The spectrum $P_{\rm e}(k)$ in Eq. (\ref{PS}) contains the shot noise contribution
\beq
 P_{\rm e,shot} = \frac{L^3}{N_{\rm eff}},
 \label{PS_shot_noise}
\eeq
where $N_{\rm eff}$ is the effective number of gas particles in the box.  
Denoting $m_{\rm i}$ as the free-electron mass of the i-th gas particle, we have $N_{\rm eff}= (\sum_{\rm i} m_{\rm i})^2/(\sum_{\rm i} m_{\rm i}^2$).
If all gas particles have equal mass, then $N_{\rm eff}=N_{\rm gas}$ (where $N_{\rm gas}$ is the number of gas particles).
The shot noise was subtracted from the measured $P_{\rm e}(k)$.
Here we measured $P_{\rm e}(k)$ up to $z=5$, because $P_{\rm e}(k)$ is noisy for $z \geq 6$ owing to the low free-electron abundance. % at such high $z$.

We also measured the matter power spectrum $P_{\rm dmo}(k)$ in the dark-matter-only (dmo) run (TNG300-1-Dark in Table\ref{table_TNG}).
This spectrum $P_{\rm dmo}(k)$ can be used to clarify the difference in clustering between free electrons and dark matter.

Figure \ref{fig_pk} plots the measured power spectra at several redshifts ($z=0$--$3$). 
The purple, blue and green symbols are $P_{\rm e}(k)$, while the grey circles are $P_{\rm dmo}(k)$. 
The dotted lines denote the shot-noise contribution.
The vertical axis is $k^2 P_{\rm e}(k)$, which represents the contribution to the DM variance per ln$k$ from Eq. (\ref{dm_var}).
The figure shows that density fluctuations at $k = 1$--$10 \, h \, \Mpc^{-1}$ (corresponding to a scale of $2 \pi/k \simeq 1 \, \Mpc$) contribute most to the DM variance.     
The shot noise is negligibly small around this peak.  
The spectrum $P_{\rm e}(k)$ agrees with $P_{\rm dmo}(k)$ at large scales ($k < 1 \, h \, \Mpc^{-1}$) but is strongly suppressed at intermediate and small scales ($k \gtrsim 1 \, h \, \Mpc^{-1}$).   
\citet{Springel2018} previously measured the power spectrum of the gas in the TNG simulations and gave a physical explanation for this suppression: the stellar and AGN feedback expels gas from the halos and suppresses the gas clustering, especially at low $z$, but gas cooling enhances clustering at very small scales, $k> 10 \, h \, \Mpc^{-1}$.
In fact, $k^2 P_{\rm e}(k)$ rises slightly for $k> 10 \, h \, \Mpc^{-1}$, especially for the higher-resolution run.
%At smallest scales ($k \gtrsim 10 \, h \, \Mpc^{-1}$), the results approach the shot noise prediction (\ref{PS_shot_noise}) especially for TNG300-3.
%At the largest scales, the free electrons trace the underlying dark-matter distribution.
The results for $P_{\rm e}(k)$ at small scales ($k \gtrsim 10 \, h \, \Mpc^{-1}$) do not converge among the different-resolution runs owing to the lack of spatial resolution. 
%The higher resolution run gives larger small-scale power.
%As the DM variance $\sigma_{\rm DM}^2$ is proportional to $\int d\ln k k^2 P_{\rm e}(k)$ in Eq.(\ref{dm_var}), 
%As the peak of $k^2 P_{\rm e}(k)$ most contributes the DM variance in Eq.(\ref{dm_var}),  
The dashed-orange curves are Halofit results from a fitting formula for non-linear $P_{\rm dmo}(k)$ \citep{Smith2003,Takahasi2012}.
These curves agree with the dark-matter-only simulation results very well. 

To model $P_{\rm e}(k;z)$, we introduce the bias factor $b_{\rm e}(k;z)$ defined by 
\beq
  b^2_{\rm e}(k;z) \equiv \frac{P_{\rm e}(k;z)}{P_{\rm dmo}(k;z)}.
  \label{bias}
\eeq
Figure \ref{fig_bias} plots the measured bias.
The bias approaches unity in the small-$k$ limit, but it is suppressed at large $k$ ($\gtrsim 1 \, h \, \Mpc^{-1}$). 
%The small-scale discrepancy among the different symbols shows the lack of spatial resolution. 
At the largest scales (i.e., the smallest $k$), the bias is very close to unity, although it is slightly smaller than unity (by approximately a few percent), especially at high $z$.\footnote{\cite{Shaw2012} previously measured $b_{\rm e}^2(k)$ from their hydrodynamic simulations. Their result is somewhat smaller than ours in the low$-k$ limit: $b_{\rm e}^2(k)=0.6$--$1$ and varies with $z$ (see the right panel of their Fig. 2).
However, according to the cosmological perturbation theory of mixed components (baryons and dark matter), the baryon-fluctuation amplitude is only slightly smaller ($<4 \%$) than the dark matter one for $k \leq 0.1 \, \, h \, \Mpc^{-1}$ and $z \leq 3$ \citep[e.g.,][their Fig.1]{Somogyi2010}.}
This is because the baryon-density fluctuations gradually catch up to the dark matter fluctuations after the epoch of decoupling (at $z \simeq 1100$).
The red curves are our fits to TNG300-1, where the bias is calibrated at $10$ redshifts in the range $z=0$--$5$ ($z=0, 0.2, 0.4, 0.7, 1, 1.5, 2, 3, 4$ and $5$).
The range of $k$ included in the fit is determined such that the TNG300-1 and -2 results agree to within $20 \, \%$.
The bias factor is fitted by the function
\beq
  b^2_{\rm e}(k;z) = \frac{b_*^2(z)}{1+\left\{ {k}/{k_*(z)} \right\}^{\gamma(z)}},
  \label{bias_fit}
\eeq
with
\begin{align}
 b_*^2(z) &= 0.971-0.013 \, z, \nonumber \\
 \gamma(z) &= 1.91-0.59 \, z +0.10 \, z^2,  \nonumber \\
 k_*(z) &= 4.36 - 3.24 \, z + 3.10 \, z^2 -0.42 \, z^3,
 \nonumber
\end{align}
where $k_*$ has units of $h \, \Mpc^{-1}$.
This function agrees with the simulation results for $P_{\rm e}(k;z)/P_{\rm dmo}(k;z)$ to within $3.5 \, (10.8) \%$ for $k<2 \, (10) \, h \Mpc^{-1}$ in the range $z=0$--$5$.

The user can compute $P_{\rm e}(k)$ from the bias factor (\ref{bias_fit}) and the $P_{\rm dmo}(k)$ model.
Accurate fitting formulas for non-linear $P_{\rm dmo}(k)$ have been presented, such as Halofit \citep{Smith2003,Takahasi2012}, HMcode \citep{Mead2015} and the Mira-Titan emulator \citep{Lawrence2017}.
% and EuclidEmulator \citep{EuclidEmu2019}.
These formulas agree with the latest dark-matter simulations to within $5 \%$ up to $k=10 \, h \, \Mpc^{-1}$ \citep[e.g.,][their Fig. 6]{SA2019}. 
Halofit and HMcode are implemented in public codes such as CAMB\footnote{\url{https://camb.info/}} and CLASS\footnote{\url{http://class-code.net/}}.

\section{Making mock sky maps of the DM}

\begin{figure*}
 \includegraphics[width=2 \columnwidth]{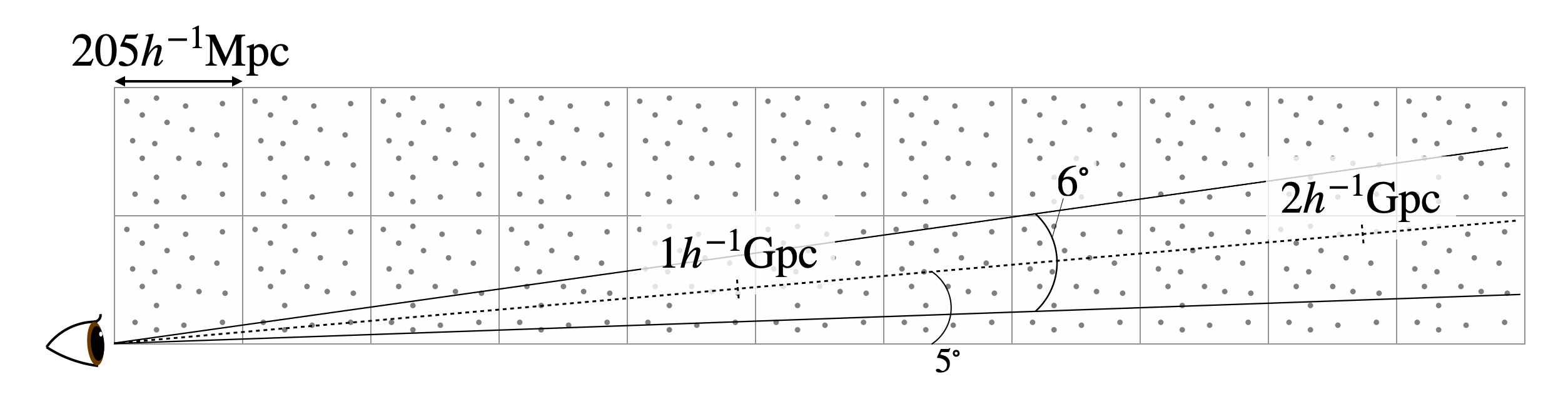}
 \caption{Schematic picture of our ray-tracing simulation setting. The grey squares represent the TNG300 simulation boxes with periodic boundary conditions. %$r$ is the comoving distance along the line-of-sight from the observer. 
The field of view is $6 \times 6 \, {\rm deg}^2$.}
 \label{fig_raytracing}
\end{figure*}

\begin{figure}
 \hspace{-2.cm}
 \includegraphics[width=1.5 \columnwidth]{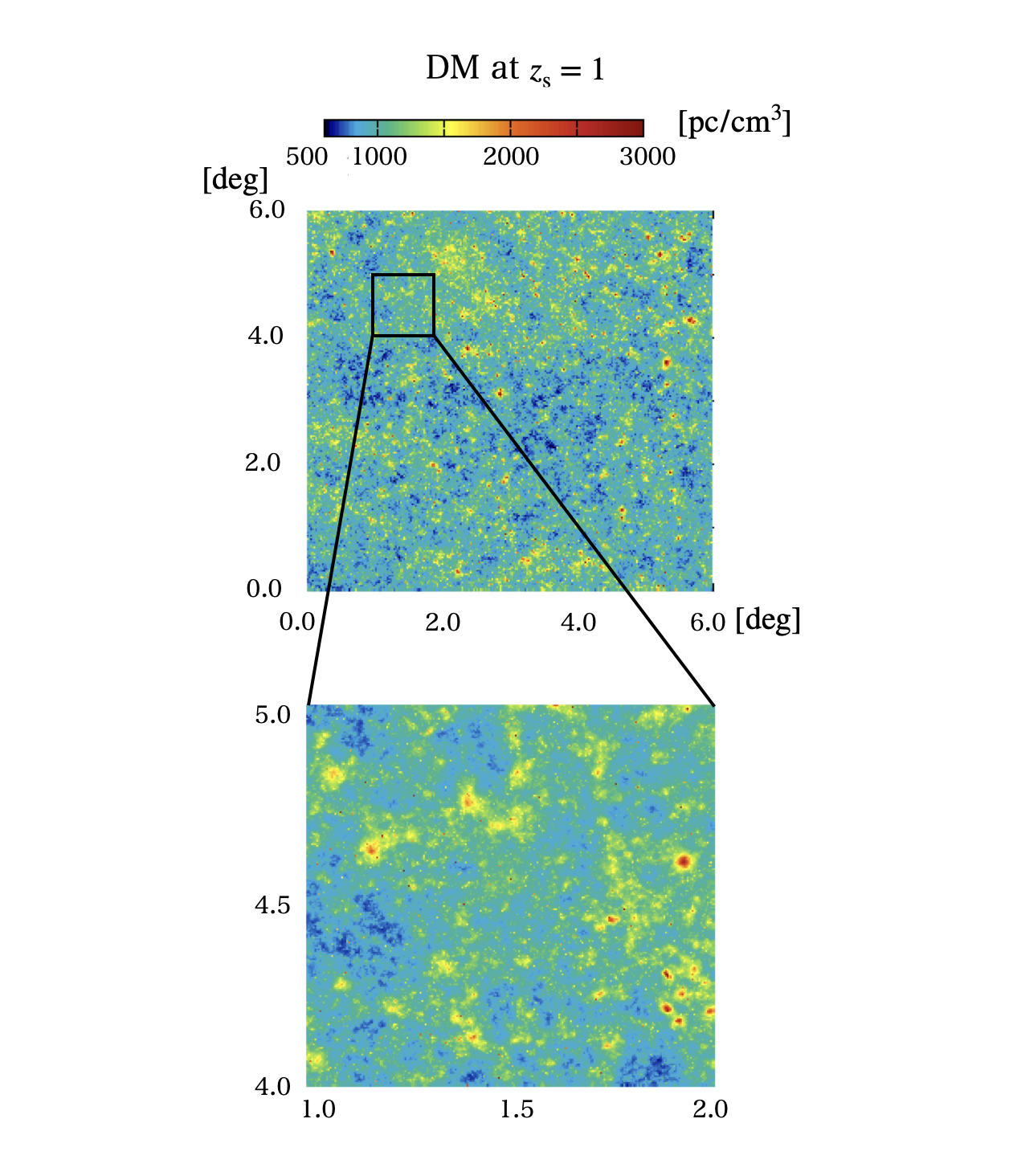}
 \vspace{-0.5cm}
 \caption{Contour plot of the DM at source redshift $z_{\rm s}=1$. The field of view is $6 \times 6 \, {\rm deg}^2$. The bottom panel is a zoom-in map of $1 \times 1 \, {\rm deg}^2$. In this and the following figures, DM refers to the cosmological dispersion measure (excluding the Milky Way and host-galaxy contributions).}
 \label{fig_dm_map}
\end{figure}

This section describes our procedure for making mock maps of the DM. 
We placed the simulation boxes along the line-of-sight direction using periodic boundary conditions, as shown in Fig. \ref{fig_raytracing}.
The observer is placed at a corner of the box at $z=0$.
The field of view was set to be a square of $6 \times 6 \, {\rm deg}^2$. 
To avoid repeating the same structure along the line of sight, we tilted the main axis (denoted by the dotted line) of the line of sight by $5$ deg from the box axis. 
%The each simulation box corresponds to a redshift (given in Table \ref{table_list}).
The lower-$z$ box in Table \ref{table_list} was placed closer to the observer.
For a given comoving distance $r$, we used the box nearest to $r$.
For instance, from Table \ref{table_list}, the lowest-$z$ box (at $z=0$) was used for $r/(h^{-1} \, \Mpc) \leq 293/2$, the second-lowest box (at $z=0.1$) was used for $293/2<r/(h^{-1} \, \Mpc) \leq (293+571)/2$ and so on.
Note that, due to the periodicity of the box, for $r > 205 \, h^{-1} \, \Mpc/(6 \, {\rm deg}) \approx 2.0 \, h^{-1} \, \Gpc$, the same structure may appear more than once in the field of view. 

Free electrons are included in the gas particles.
For each gas particle, the TNG team provided the spatial position, gas mass $m_{\rm gas}$, density $\rho_{\rm gas}$ and free-electron number density $n_{\rm e}^{\rm gas}$.
We assume that each gas particle is described by a sphere of constant density with the radius $r_{\rm gas}$ determined via $m_{\rm gas}=(4 \pi r_{\rm gas}^3/3) \rho_{\rm gas}$.
%The mean radius measured from all gas particles are given in Table \ref{table_TNG}.
%At a lower redshift, the mean radius decrease but its variance increases.   

The DM is rewritten from Eq. (\ref{dm}) as
\beq
{\rm DM}(\bftheta;z_{\rm s}) = \int_{0}^{r_{\rm s}} \!\!\! dr  n_{\rm e} (\bfr;z) \left( 1+z(r) \right),
\label{dm2}
\eeq
where $r_{\rm s} \equiv r(z_{\rm s})$ is the comoving distance to the source,
%, $\bftheta = (\theta_1,\theta_2)$ is an angular direction vector pointing to $\bfr$, and $r$ and $n_{\rm e}$ are given in the comoving unit.
and $\bftheta$ denotes the angular position in the field of view, i.e., $\bftheta=(\theta_1,\theta_2)$ with $|\theta_{1,2}| \leq 3$ deg.
Light rays are emitted from the observer and propagate along straight lines in the field. 
The DM is computed by summing the contributions from all gas particles intersecting the light-ray path:
\beq
{\rm DM}(\bftheta;z_{\rm s}) = \sum_{\rm i} n_{\rm e,i}^{\rm 2D}(b_{\rm i}) (1+z_{\rm i}),
\label{dm_sum}
\eeq
where $n_{\rm e,i}^{\rm 2D}$ is the free-electron column density of the i-th gas particle and $b_{\rm i}$ is the impact parameter (i.e., the minimum separation between the ray path and the position of the i-th particle).
The redshift $z_{\rm i}$ is calculated from the comoving distance using Eq. (\ref{r-z_relation}). 
Assuming that the i-th gas particle has a constant density $n_{\rm e,i}^{\rm gas}$ and radius $r_{\rm gas,i}$, its column-density profile is given by
\beq
  n_{\rm e,i}^{\rm 2D}(b_{\rm i}) = 2 \, n_{\rm e,i}^{\rm gas} \sqrt{\,r_{\rm gas,i}^2-b_{\rm i}^2} \, ,
\eeq
for $b_{\rm i}<r_{\rm gas,i}$, and $n_{\rm e,i}^{\rm 2D}(b_{\rm i}) = 0$ otherwise.
The DM in Eq. (\ref{dm_sum}) is computed along the straight-line ray-path up to $z_{\rm s}=3$.
We comment that each light ray passes through a sufficient number of gas spheres.
For instance, in TNG300-1, there are $1.0 \times 10^4$, $2.2 \times 10^4$ and $3.7 \times 10^4$ gas spheres intersecting a single light ray up to $z_{\rm s}=0.4$, $1$ and $2$, respectively.
For TNG300-2 (-3), this number simply decreases by a factor of $2$ $(4)$. 
If the number of intersecting gas spheres follows a Poisson distribution, the accuracy of the DM in Eq. (\ref{dm_sum}) is roughly given by $({\rm the~number})^{-1/2}$.

We homogeneously emitted $5400^2$ rays through the $6 \times 6 \, {\rm deg}^2$ field and computed their DMs using Eq. (\ref{dm_sum}).
The resulting angular resolution is $4$ arcsec ($=6 \, {\rm deg}/5400$).   
We stored the DM data up to $z_{\rm s}=3$ at every $\Delta z_{\rm s}=0.02$ step.
To see the statistical variation among the maps, we prepared $10$ such maps\footnote{This number 10 is limited by our hard-disk storage, but sufficient for our studies.} by recycling the same simulation data.
Here the recycling procedure is as follows: (i) swap the coordinates (e.g., $x \leftrightarrow y$) for all particles in the box,  (ii) shift the coordinates (e.g., $x \rightarrow x+x_0$ with an arbitrary constant $x_0$ where the coordinate origin can be freely chosen under the periodic boundary conditions) for all particles and (iii) finally place these boxes as in Fig. \ref{fig_raytracing} and perform the same ray-tracing calculation.
The swapped coordinates (i) and the coordinate shift (ii) were randomly chosen for each map. 
We prepared the 10 maps for each of the three resolution runs. % (from TNG300-1 to -3).
We checked that the observer does not belong to any halo (the TNG also provides halo catalogues containing halo positions and radii), and thus, the measured DM does not contain the observer's halo contribution. 

Figure \ref{fig_dm_map} is a contour map of the DM from TNG300-1 at $z_{\rm s}=1$. 
This is one of the $10$ maps.
The red (blue) regions correspond to foreground clusters or galaxies (voids). 
We present an analysis of the $10$ maps in the following section.

\section{Results}

This section presents measurements of the DM statistics from the mock maps: the mean and variance (subsection 5.1), probability distribution of the DM (subsection 5.2), probability distribution of $z_{\rm s}$ for a given DM (subsection 5.3), angular power spectrum (subsection 5.4) and angular correlation function (subsection 5.5).
Comparisons with the analytical results using the fitting functions (given in section 3) are also presented. 

\subsection{Mean and variance of the DM}

\begin{figure}
 %\hspace{-1.2cm}
 %\includegraphics[width=1.15 \columnwidth]{/Users/takahasi/nc/frb_cosmology/variance_DM/figs/fig_DM_mean_rms_paper.pdf}
 \includegraphics[width=1.15 \columnwidth]{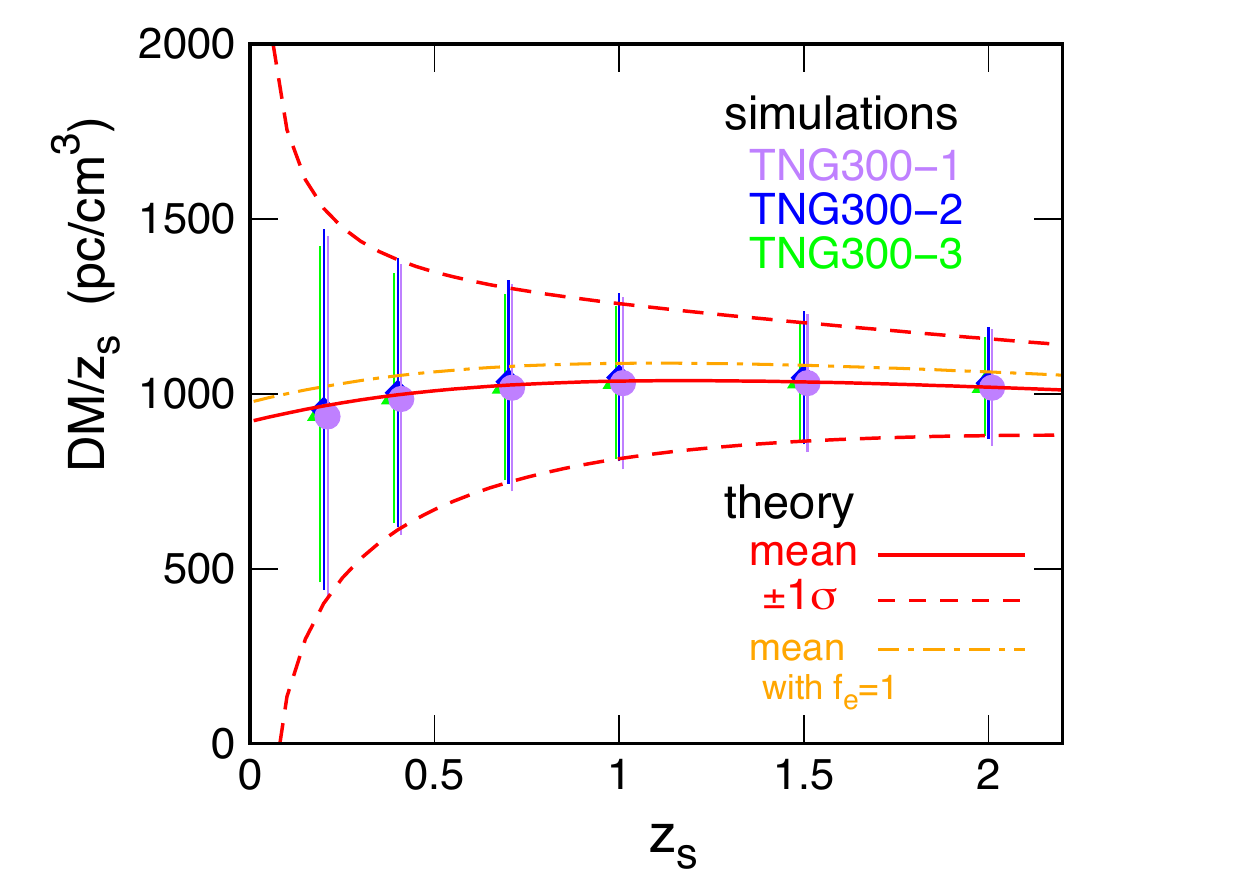}
 %\vspace{-0.5cm}
 \caption{Mean with $1 \sigma$ standard deviation of the DM as a function of source redshift $z_{\rm s}$. The purple, blue and green symbols are the simulation results measured from the $10$ mock maps (one of them is plotted in Fig. \ref{fig_dm_map}). The solid (dashed) red curve denotes the analytical predictions for the mean (standard deviation) discussed in section 2. The dash-dotted orange curve is the same as the solid one, but assuming $f_{\rm e}=1$. In this and following figures, the theory refers to the analytical model (in section 2) using the fitting formulas for $f_{\rm e}(z)$ and $P_{\rm e}(k;z)$ (in section 3). }
 \label{fig_dm_mean_var}
\end{figure}

\begin{figure}
 %\hspace{-1.2cm}
 %\includegraphics[width=1.15 \columnwidth]{/Users/takahasi/nc/frb_cosmology/variance_DM/figs/fig_DM_rms_paper.pdf}
 \includegraphics[width=1.15 \columnwidth]{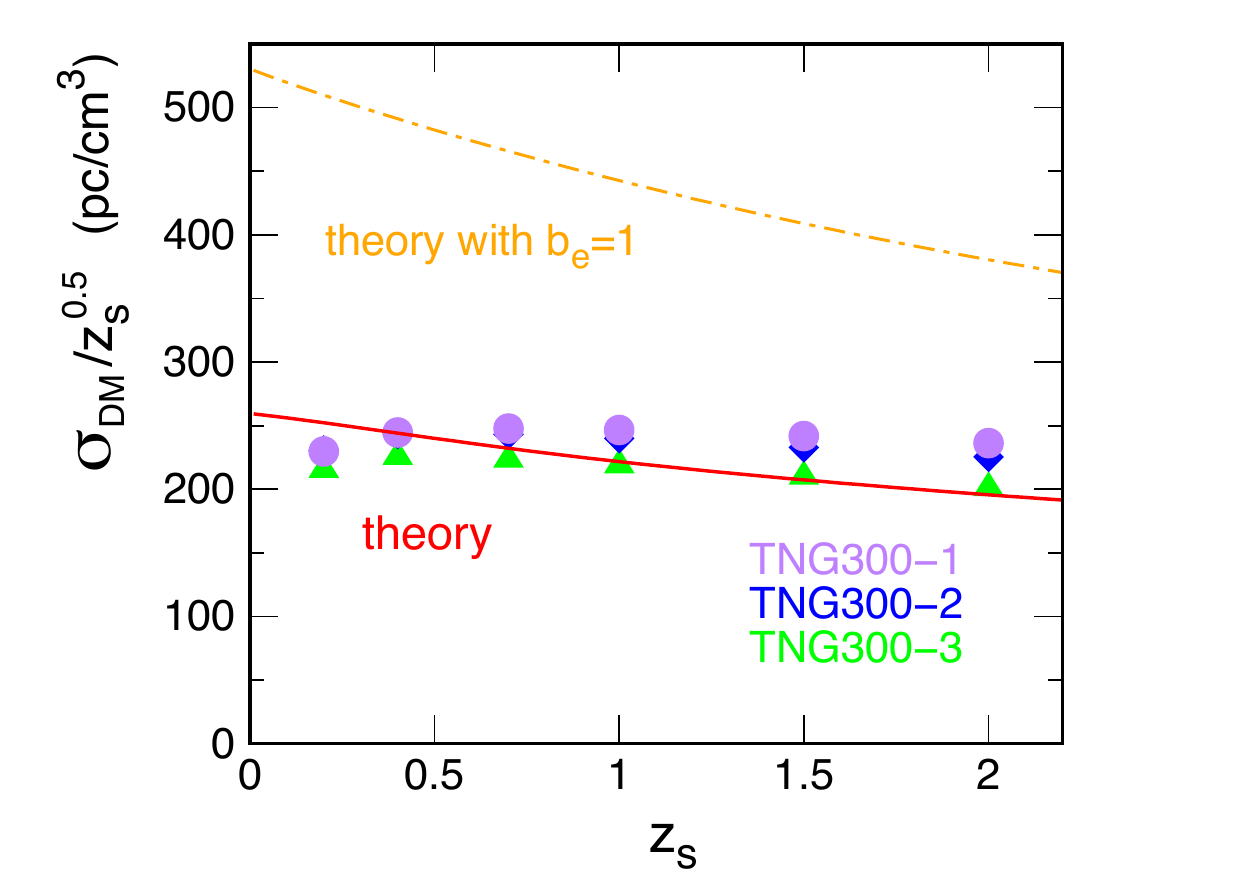}
 %\vspace{-0.5cm}
 \caption{Standard deviation of the DM as a function of $z_{\rm s}$. The purple, blue and green symbols are the TNG300-1, -2 and -3 results. The solid red curve denotes the analytical prediction given in Eq. (\ref{dm_var}). The dash-dotted orange curve is the same as the solid one, but assuming that the free electrons exactly trace the dark matter (i.e., $b_{\rm e}=1$).}
 \label{fig_dm_var}
\end{figure}

We measured the mean and variance of the DM from the $10$ mock maps. 
As there are $5400^2$ data points in each map, the total number of rays ($=10 \times 5400^2 \simeq 2.9 \times 10^8$) is sufficient for statistical analysis.
Figure \ref{fig_dm_mean_var} plots the mean with the standard deviation as a function of $z_{\rm s}$.
The simulation results are measured at several values of $z_{\rm s}$ ($=0.2, 0.4, 0.7, 1, 1.5$ and $2$).
The mean and variance are obtained by using all the rays in the $10$ maps. % (not in each map).
The solid and dashed red curves are the analytical mean and standard deviation given in Eqs. (\ref{mean_DM}) and (\ref{dm_var}), respectively.  
Here and hereafter, the fitting formulas for $f_{\rm e}(z)$ and $P_{\rm e}(k;z)$ given in section 3 are used to compute the analytical predictions, and the theory refers to the analytical model (in section 2) using these fitting formulas. 
The figure shows that the theory agrees with the simulations very well.
The mean DM is approximately proportional to $z_{\rm s}$, $\overline{\rm DM}(z_{\rm s}) \approx 1000 \times z_{\rm s} \, {\rm pc} \, {\rm cm}^{-3}$, which is consistent with previous work \citep[e.g.,][]{Ioka2003}.
The dash-dotted orange curve is the mean DM for the fully ionised case ($f_{\rm e}=1$), as assumed in previous studies \citep{Ioka2003,Inoue2004}. 
This simple assumption only slightly overestimates the mean by approximately a few percent in this redshift range.
%(the deviation is within the error bars).

Figure \ref{fig_dm_var} is similar to Figure \ref{fig_dm_mean_var} but plots only the standard deviations.
The agreement between the theory and the simulations is within $10$--$20 \, \%$.
At large $z_{\rm s} \, (\gtrsim 1)$, the theory gives slightly lower values than the simulations.
This is because $P_{\rm e}(k)$ was fitted up to $k=10 \, h \, \Mpc^{-1}$, and gas cooling slightly enhances $P_{\rm e}(k)$ for $k>10 \, h \, \Mpc^{-1}$ (see Fig. \ref{fig_pk}), so fluctuations smaller than this fitting range provides additional contributions to $\sigma_{\rm DM}$, especially at high $z_{\rm s}$.
%We comment that at small $z_{\rm s} (\lesssim 0.2)$, the $\sigma_{\rm DM}$ has large scatter among the maps ($\sim ... \%$ level), because the late-time density fluctuations are getting non-Gaussian.
The TNG300-1 and -2 results almost converge, because fluctuations with $k=1$--$10 \, h \, \Mpc^{-1}$ contribute most to $\sigma_{\rm DM}$ (see also subsection 3.3), and these runs resolve this scale sufficiently.
The TNG300-3 results give a slightly smaller result because of the lowest resolution.    
From this figure, the standard deviation is roughly given by $\sigma_{\rm DM}(z_{\rm s}) \approx 230 \times z_{\rm s}^{0.5} \, {\rm pc} \, {\rm cm}^{-3}$, which is consistent with the halo-model prediction \citep[][see also \cite{KL2019}] %who gave a relation $\sigma_{\rm DM}(z_{\rm s}) \approx 210 \, z_{\rm s}^{0.5}$. ]{
{McQuinn2014,MacQ2020}.
The previous ray-tracing simulation \citep[][see the dashed line in their Fig. 1]{Jaros2019} gave $\sigma_{\rm DM} \simeq 200$ and $300 \, {\rm pc} \, {\rm cm}^{-3}$ at $z_{\rm s}=1$ and $2$, respectively, which are also consistent with our results. 
%\rtrv{We comment that the analytical result for $\sigma_{\rm DM}$ is proportional to $f_{\rm e}$ from Eq. (\ref{dm_var}).}

Previously, \cite{Shirasaki2017} and \citet{Pol2019} studied the DM statistics using dark-matter only simulations, assuming the free-electron distribution to be the same as the dark matter distribution (i.e., $b_{\rm e}=1$).  
The dash-dotted orange curve in Fig. \ref{fig_dm_var} corresponds to this case.
This assumption may overestimate $\sigma_{\rm DM}$ by a factor of two.

We comment that although the TNG300 simulations do not contain density fluctuations larger than the box size $L \, (=205 \, h^{-1} \, {\rm Mpc})$, this does not affect the results for $\sigma_{\rm DM}$.
%we confirmed that such larger-scale fluctuations does not affect $\sigma_{\rm DM}$.
If we set the large-scale cutoff $P_{\rm e}(k)=0$ for $k<2 \pi /L \simeq 0.03 \, h \, \Mpc^{-1}$ in Eq. (\ref{dm_var}), the variance $\sigma_{\rm DM}^2$ is underestimated by $<0.5 \, \%$ in the range $z_{\rm s}=0$--$2$, because $k^2 P_{\rm e}(k)$ at such a large scale is too small to give a contribution.
%the $\sigma_{\rm DM}^2$ is very sensitive to the nearby structure.
%In fact, in the limit of $r \rightarrow 0$, the integrand of Eq. (\ref{dm_var}) is $P(k=\ell/r)/r^2 \propto r^{\alpha-2}$ with $\alpha \approx 5$.

\subsection{Probability distribution of the DM}

\begin{figure*}
 \vspace{-2.cm}
 \includegraphics[width=2. \columnwidth]{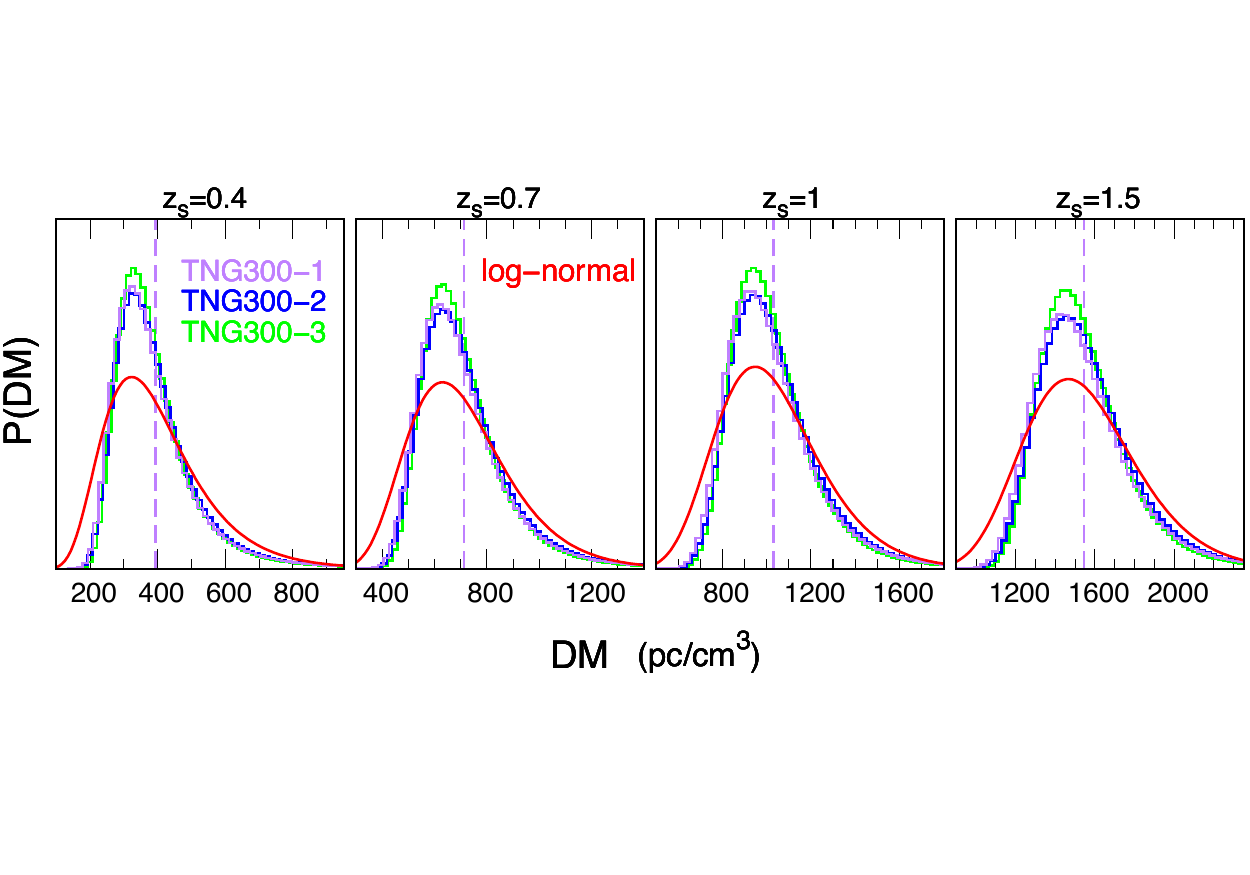}
 \vspace{-2.5cm}
 \caption{Probability distributions of the DM for several source redshifts $z_{\rm s}$, as measured from the mock maps. The purple, blue and green histograms are the TNG300-1, -2 and -3 results. The vertical dashed line denotes the mean DM measured from TNG300-1. The red curves are log-normal distributions. The vertical axis is in arbitrary units.}
 \label{fig_pdf-dm}
\end{figure*}

\begin{figure}
 %\hspace{-1.2cm}
 %\includegraphics[width=1.1 \columnwidth]{/Users/takahasi/nc/frb_cosmology/pdf/figs/fig_pdf-dm_fitform_z0.5_paper.pdf}
 \includegraphics[width=1.1 \columnwidth]{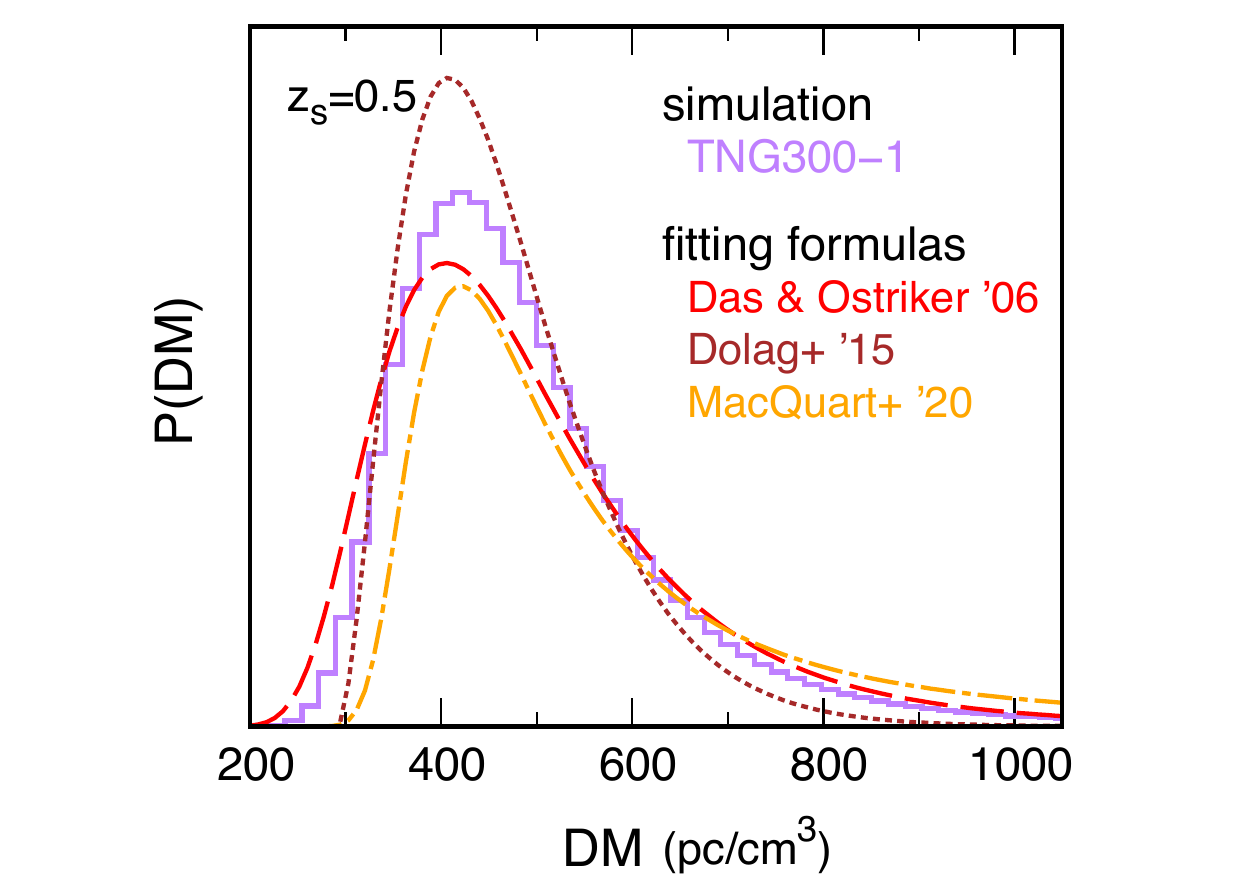}
 %\vspace{-0.5cm}
 \caption{Comparison of probability distributions of the DM at $z_{\rm s}=0.5$. The purple histogram is the TNG300-1 result, while the curves denote previous fitting formulas: \citet[][dashed red]{Das2006}, \citet[][dotted brown]{Dolag2015} 
 %\citet[][dotted gray]{Walters2019}, 
 and \citet[][dash-dotted orange]{MacQ2020} in subsection 5.2.}
 \label{fig_pdf-dm_fitform}
\end{figure}

Figure \ref{fig_pdf-dm} plots the probability distribution function (PDF) of the DM for several source redshifts ($z_{\rm s}=0.4, 0.7, 1$ and $1.5$).
The coloured histograms correspond to the different TNG resolutions, which are consistent with each other. 
The PDF is highly skewed, especially at low $z_{\rm s}$, owing to the strong non-Gaussianity of the density fluctuations.
The red curves are log-normal distributions with the mean and variance given by the DM-map measurements from TNG300-1.
At higher $z_{\rm s}$, the simulations approach log-normal distributions.
This model is roughly consistent with the simulations, but it has broader tails around the peak and is less skewed than the simulations. 
%The vertical dashed lines denote the mean DM in TNG300-1.

Figure \ref{fig_pdf-dm_fitform} shows a comparison of the PDF with previous fitting formulas.
\cite{Das2006} measured the PDF of the projected matter density using dark-matter N-body simulations. 
Their purpose was to investigate the PDF of the weak-lensing convergence field. 
They proposed a modified log-normal distribution (given in their Eq. (11)). 
\cite{Dolag2015} performed cosmological hydrodynamic simulations and measured the PDF. 
Their formula depends only on $z_{\rm s}$ (given in their Eq. (6)).
%\footnote{Here we normalized their PDF by ourselves, instead of using their proposed normalization. Because it gave a too small amplitude.}     
%Their formula of the amplitude, in their Eq. (7), is too small.}.
%\cite{Walters2019} calculated the DM statistics by intervening galactic halos using the halo-model approach \citep{PZ2019}. 
%They provided a slightly modified lognormal model.  
\cite{MacQ2020} proposed a skewed Gaussian PDF calibrated by the halo-model prediction \citep{McQuinn2014}. Here we used their best-fit model (their Eq. (4) with $\alpha=\beta=3$).
In \cite{Das2006} and \cite{MacQ2020}, the formulas contain free parameters, but they are fully determined by the mean and variance of the DM measured from TNG300-1.   
%The \cite{Walters2019} is very similar to the standard lognormal.
The figure shows that all the formulas show better agreement with the simulation than the simple log-normal model.
%The simulation is almost consistent with all of them.

\subsection{Probability distribution of the source redshift for a given DM}

\begin{figure}
 %\hspace{-1.2cm}
 %\includegraphics[width=1.15 \columnwidth]{/Users/takahasi/nc/frb_cosmology/pdf/figs/fig_z-dm_mean_paper.pdf}
 \includegraphics[width=1.15 \columnwidth]{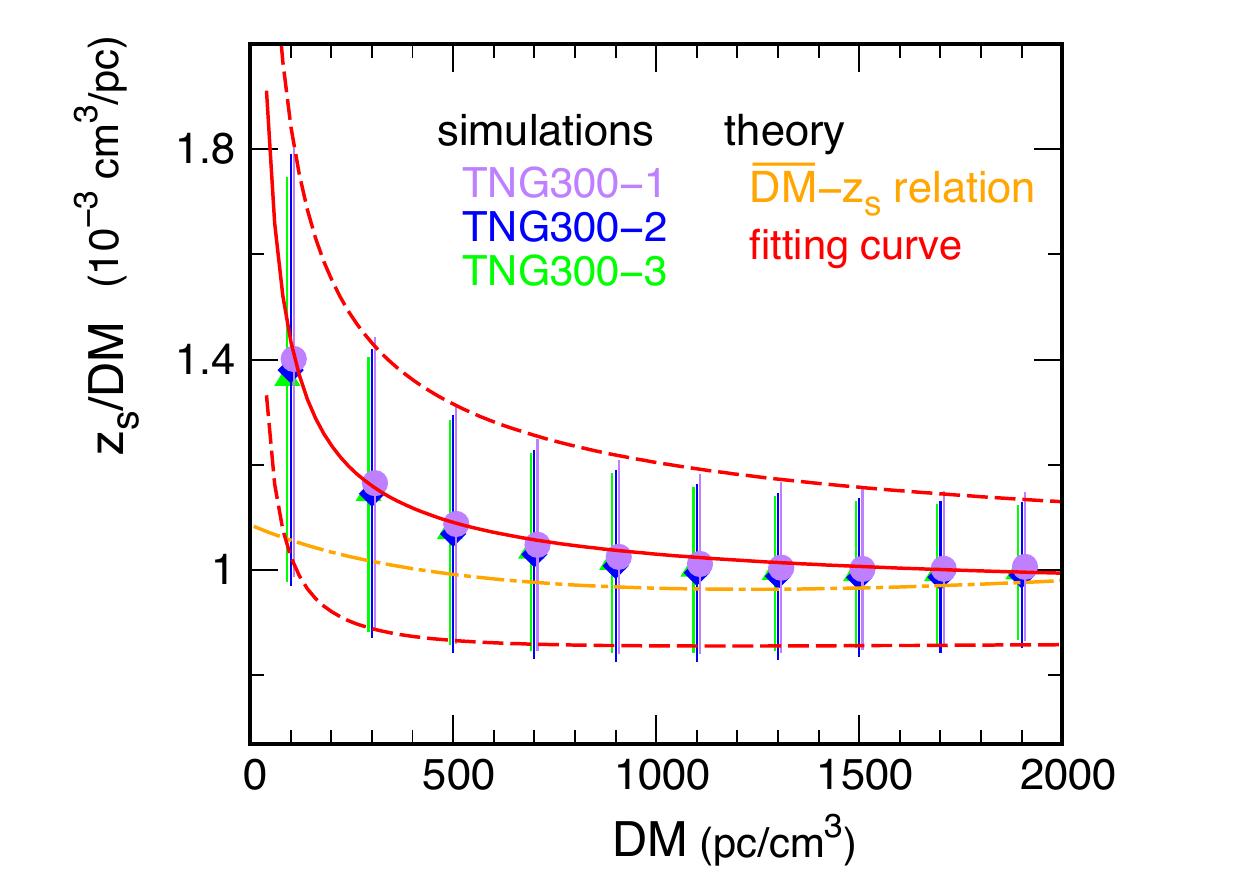}
 %\vspace{-0.5cm}
 \caption{Source redshift inferred from the measured DM. The symbols with $1 \sigma$ error bars are the simulation results. The dash-dotted orange curve is the analytical $\overline{\rm DM}$-$z_{\rm s}$ relation (\ref{mean_DM}). The solid (dashed) red curve is the fit to the mean ($1 \sigma$ error bars), given in Eq. (\ref{z-DM_fit}).}
 \label{fig_z-dm_mean}
\end{figure}

\begin{figure*}
 \vspace{-2.cm}
 \includegraphics[width=2. \columnwidth]{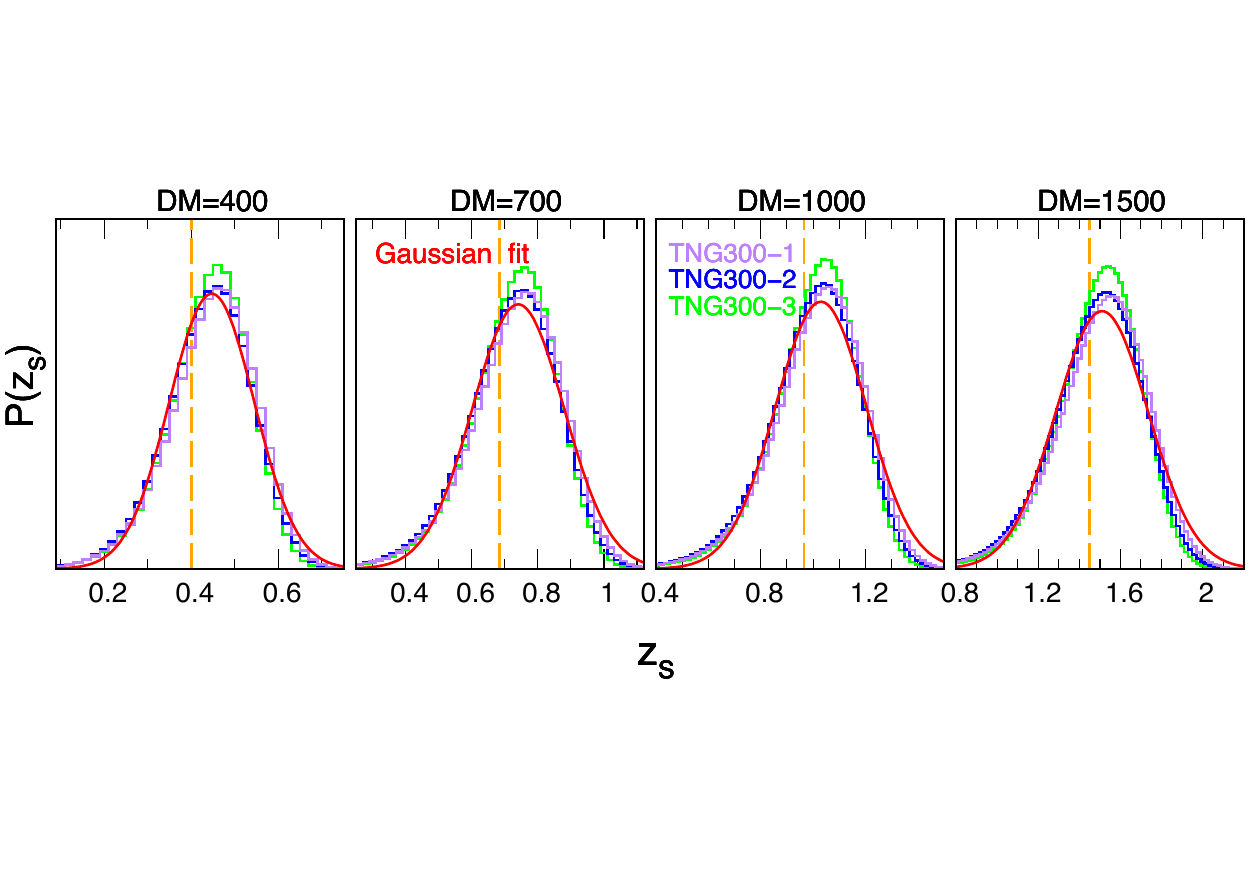}
 \vspace{-2.5cm}
 \caption{Probability distributions of the source redshift for several given DMs (as denoted over each panel, in units of ${\rm pc \, cm^{-3}}$), as measured from the mock maps. The purple, blue and green histograms are the TNG300-1, -2 and -3 results. The vertical dashed-orange lines are the analytical $\overline{\rm DM}$--$z_{\rm s}$ relation (\ref{mean_DM}). The red curves are Gaussian fits with mean and variance given by Eq. (\ref{z-DM_fit}). The vertical axis is in arbitrary units.}
 \label{fig_pdf-z}
\end{figure*}

%This subsection discusses the statistics of source redshift $z_{\rm s}$ inferred from the measured DM.
\rtrv{Since the DM was calculated at every source redshift step $\Delta z_{\rm s}=0.02$ up to $z_{\rm s}=3$ for each light ray (see section 4), we can obtain the source redshift $z_{\rm s}$ corresponding to a given DM. Using all the rays ($=2.9 \times 10^8$),  the distribution of $z_{\rm s}$ for a given DM is also obtained.} % using all the rays ($=2.9 \times 10^8$).}
Figure \ref{fig_z-dm_mean} plots the mean and $1 \sigma$ standard deviations for the source redshift $z_{\rm s}$ inferred from the measured DM.
The coloured symbols are the simulation results, while the dash-dotted orange curve is the analytical $\overline{\rm DM}$--$z_{\rm  s}$ relation (\ref{mean_DM}). 
As clearly shown in the figure, the $\overline{\rm DM}$--$z_{\rm s}$ relation underestimates $z_{\rm s}$, especially at low $z_{\rm s}$, owing to the highly skewed distribution of the DM (shown in Figs. \ref{fig_pdf-dm} and \ref{fig_pdf-dm_fitform}). 
As the peak of the DM is \textit{lower} than the analytical mean for a given $z_{\rm s}$ in Fig. \ref{fig_pdf-dm}, the inferred $z_{\rm s}$ is \textit{higher} than the analytical mean for a given DM.
This trend is consistent with a previous finding \cite[][their Fig. 3]{Pol2019}. 
The figure shows that the standard deviation of $z_{\rm s}$ is approximately $20 \, \%$ for ${\rm DM} > 500 \, {\rm pc \, cm}^{-3}$ but becomes larger for a nearer source.  
As the statistics of $z_{\rm s}$ are useful in searching for the host galaxy of an FRB from the measured DM, we fitted the mean $\bar{z}_{\rm s}$ and standard deviation $\sigma_{z_{\rm s}}$ from TNG300-1 in the range ${\rm DM}=100$--$2000 \, {\rm pc \, cm}^{-3}$:
\begin{align}
 \bar{z}_{\rm s}({\rm DM}) &= 0.015 \, {\rm DM}^{0.26}+9.4 \times 10^{-4} \, {\rm DM},   \nonumber  \\
 \sigma_{{z}_{\rm s}}({\rm DM}) &= 0.0024 \, {\rm DM}^{0.61}+1.2 \times 10^{-5} \, {\rm DM}, 
\label{z-DM_fit}
\end{align}
where DM is in units of ${\rm pc} \, {\rm cm}^{-3}$.
These formulas are plotted as the solid and dashed red curves in Fig. \ref{fig_z-dm_mean}.
Though the relation (\ref{z-DM_fit}) was derived from TNG300-1 for a specific $f_{\rm e}(z)$ model,  it may be used for any $f_{\rm e}(z)$ model,
%using an approximate relation $\bar{z}_{\rm s} \propto f_{\rm e}^{-1}$.
so long as $f_{\rm e}(z)$ does not evolve strongly with time (which is valid in our case at $z<2$, where $f_{\rm e} \simeq 0.95$, as shown in Fig. \ref{fig_mean_ne}).
In this case, %denoting $\tilde{f}_{\rm e}$ as an alternative model, 
the relation (\ref{z-DM_fit}) may be used by replacing ${\rm DM}$ with ${\rm DM} \times (0.95/f_{\rm e})$ for an arbitrary $f_{\rm e}$.

Figure \ref{fig_pdf-z} plots the PDFs of the $z_{\rm s}$ for given DMs.
The histograms are the simulation results, while the vertical dashed-orange lines are the values of $z_{\rm s}$ inferred from the $\overline{\rm DM}$--$z_{\rm  s}$ relation (\ref{mean_DM}).
%As seen in the previous figure \ref{fig_z-dm_mean}, 
The analytical expectation of $\bar{z}_{\rm s}$ is thus systematically lower than the true value by $10$--$20 \, \%$, especially for a low DM.
The red curves are Gaussian distributions with mean and variance given by Eq. (\ref{z-DM_fit}).
The PDF of the $z_{\rm s}$ is well described by a Gaussian.

%The figure demonstrates that $z_{\rm s}$ from the mean DM-$z_{\rm s}$ relation 

\cite{Walker2020} recently derived a PDF of the $z_{\rm s}$ by a different approach.
Their PDF is based on Bayes' theorem and uses their DM probability distribution and a given FRB redshift distribution.
Their PDF (in their Fig. 5) seems consistent with ours, but their result depends on the prior FRB redshift distribution. 
\cite{Hack2020} performed a similar analysis using the same approach.

\subsection{Angular power spectrum of the DM}

\begin{figure}
 \hspace{-1.2cm}
 \includegraphics[width=1.4 \columnwidth]{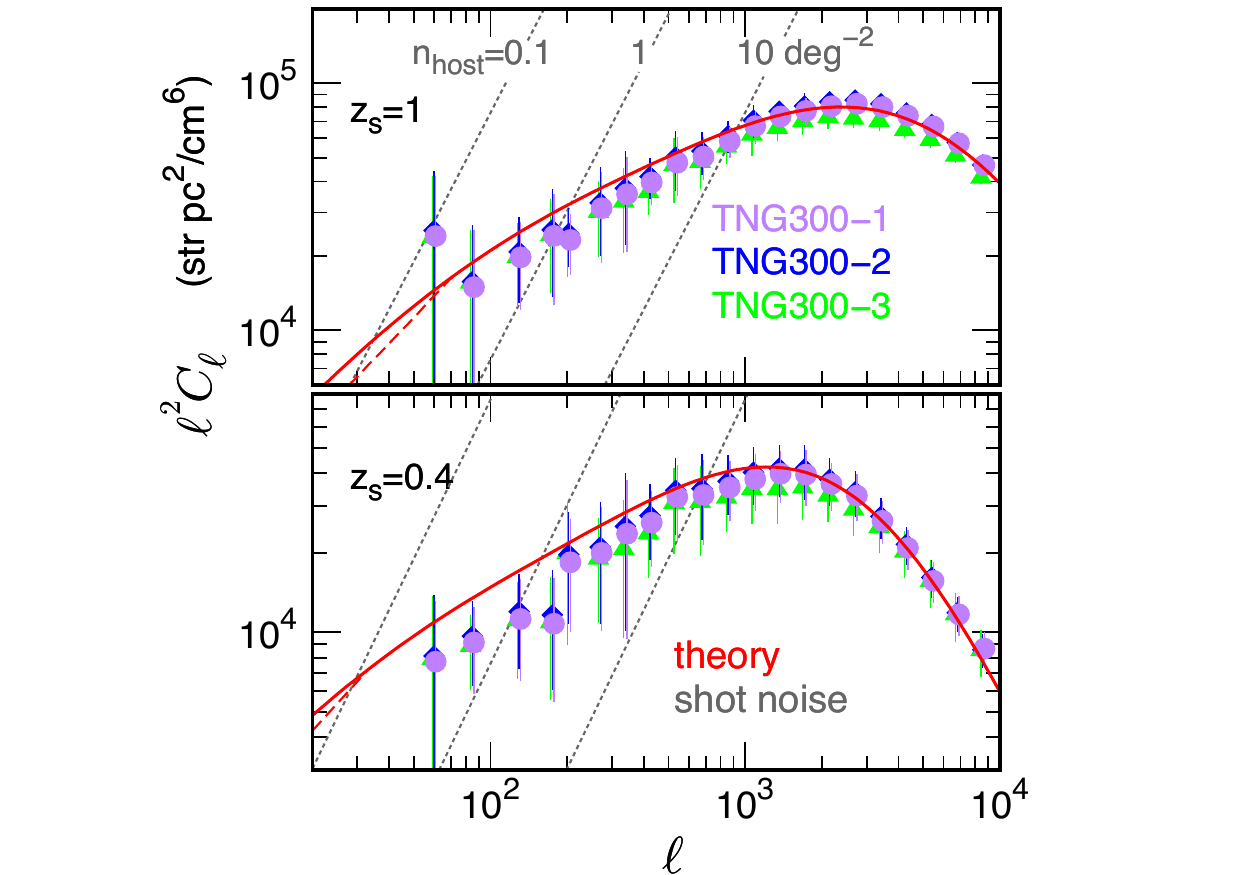}
 %\vspace{-0.5cm}
 \caption{Angular power spectra of the DM at $z_{\rm s}=1$ (upper panel) and $0.4$ (lower panel). The angular separation $\theta$ corresponding to the multipole $\ell$ is $\theta \sim \pi/\ell =  1 \, {\rm deg} (\ell/180)^{-1}$. The purple, blue and green symbols are the TNG300-1, -2 and -3 results, respectively.  The mean and $1 \sigma$ error bars (= standard deviations) are measured from the $10$ mock maps. The field of view is $6 \times 6 \, {\rm deg}^2$, and the error bars scale as $[({\rm survey \, area})/(36 \, {\rm deg}^2)]^{-1/2}$.  The solid red curves are the analytical predictions (\ref{cl_DM}). The dashed red curves are the same as the solid ones, but they include the effect of the finite size of the simulation box. The dotted grey lines denote shot noise from the host galaxies ($\sigma^2_{\rm DM,host}/n_{\rm host}$) with the intrinsic scatter of ${\rm DM}_{\rm host}$, $\sigma_{\rm DM,host}=50 \, {\rm pc} \, {\rm cm}^{-3}$, and the number densities $n_{\rm host}=0.1, 1$ and $10 \, {\rm deg}^{-2}$ from left to right.}
 \label{fig_cl}
\end{figure}

\begin{figure}
 \hspace{-0.4cm}
 \includegraphics[width=1.2 \columnwidth]{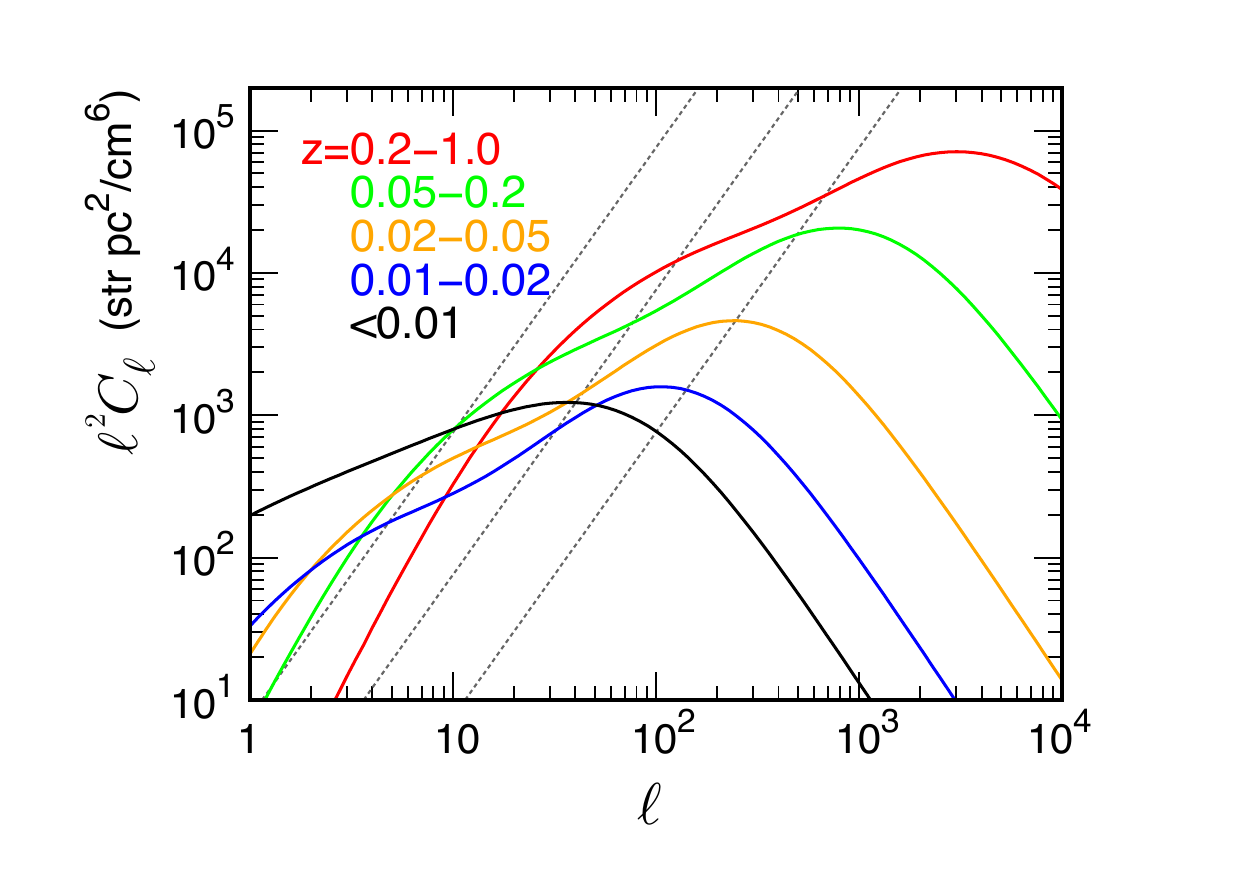}
  %\vspace{-0.5cm}
 \caption{Contribution from different redshifts to the angular power spectrum of the DM. The dotted lines are the same shot noise as in Fig. \ref{fig_cl}.}
 \label{fig_cl_diffz}
\end{figure}

We measured the angular power spectrum of the DM in the same way as discussed for the free-electron power spectrum in subsection 3.3.
The Fourier transform of the DM fluctuations in the i-th map (${\rm i}=1,2,\cdots,10$) is denoted as $\widetilde{\delta {\rm DM}}_{\rm i}(\bfell;z_{\rm s})$ from Eq. (\ref{DM_fft}).  %$\delta {\rm DM}_{\rm i}(\bftheta;z_{\rm s})$, 
Then the power spectrum for this map is obtained as
\beq
  C_{{\ell},{\rm i}}(z_{\rm s}) = \frac{1}{N_{{\rm mode},\ell}} \sum_{|\, \bfell^\prime| \in \ell} \left| \, \widetilde{\delta {\rm DM}}_{\rm i}({\bfell}^\prime;z_{\rm s}) \right|^2,
\eeq
where the summation is performed in the annulus $\ell-\Delta \ell/2<|\bfell^\prime|<\ell+\Delta \ell/2$, and $N_{{\rm mode},\ell}$ is the number of Fourier modes in the annulus with bin-width ($\Delta \log_{10} \ell=0.1$). 
We measured $C_{\ell,{\rm i}}(z_{\rm s})$ for the $10$ maps to calculate its mean and variance among the maps.

Figure \ref{fig_cl} shows the angular power spectra of the DM at $z_{\rm s}=0.4$ and $1$.
The symbols with error bars denote the simulation results for the mean and standard deviation among the maps.  
Here, the minimum multipole is determined by the side length of the map: $\ell_{\rm min}= 2 \pi/(6 \, {\rm deg}) = 60$.
The angular resolution of the maps ($= 4 \, {\rm arcsec} = 6 \, {\rm deg}/5400$) is good enough to resolve the signal up to $\ell=10^4$. 
If $C_\ell$ obeys a Gaussian distribution, the error bars scale in proportion to $[({\rm survey \, area}) \Delta \ell]^{-1/2}$, where $\Delta \ell$ is the bin-width. 
The solid red curves are the theory (\ref{cl_DM}).
The dashed red curves include the effect of the finite size of the simulation box, where we simply set $P_{\rm e}(k)=0$ for $k<2 \pi/L$ in Eq. (\ref{cl_DM}).
This effect only slightly suppresses $C_\ell$ at large scales ($\ell < 100$).
The theory and simulations agree well over a wide range of $\ell$, but the simulations are slightly suppressed at small $\ell \, (< \! 10^3)$. 
This may be caused by the sample variance of the $10$ maps (in other words, $10$ maps may not be a sufficient number to measure the mean of $C_\ell$ precisely).
\rtrv{We comment that the power spectra $C_\ell$ at different $\ell \, (\gtrsim 10^2)$ are correlated due to the non-Gaussian free-electron fluctuations \citep[the non-Gaussian covariance between different $\ell$ for the weak-lensing power spectrum was discussed in e.g.,][]{Sato2009}. The non-Gaussianity is more important for larger $\ell$ or lower $z_{\rm s}$.} %because the smaller-$\ell$ signal comes from nearby strutcures (see Fig. \ref{fig_cl_diffz}).}
The peak of $\ell^2 C_\ell$ in this figure roughly corresponds to the peak of $k^2 P_{\rm e}(k)$ in Fig. \ref{fig_pk} via $\ell_{\rm peak} \simeq k_{\rm peak} r = 2000 \, [k_{\rm peak}/(2 \, h \, \Mpc^{-1})] \, [r/(1 \, h^{-1} \, \Gpc)]$ from Eq. (\ref{cl_DM}).

In actual measurements of $C_\ell$, as indicated by several authors \citep[e.g.,][]{Shirasaki2017,Madh2019}, $C_\ell$ at small scales is strongly contaminated by shot noise from the host-galaxy contribution, ${\rm DM}_{\rm host}$.
The shot noise is given by
\beq
  C_{\rm shot}=\frac{\sigma_{\rm DM,host}^2}{n_{\rm host}},
\eeq
where $\sigma_{\rm DM,host}$ is the intrinsic scatter of ${\rm DM}_{\rm host}$, and $n_{\rm host}$ is the surface number density per steradian. 
The shot noise is plotted in Fig. \ref{fig_cl} for the cases $\sigma_{\rm DM,host}=50 \, {\rm pc \, cm}^{-3}$ and $n_{\rm host}=0.1,1$ and $10 \, {\rm deg}^{-2}$ as illustrative examples.
Roughly, the signal must exceed the shot noise in order to be detectable.
The figure shows that the small-scale signals are difficult to detect, which is consistent with the previous indication.

Figure \ref{fig_cl_diffz} plots the contribution from different redshifts to $C_\ell$. 
At smaller (larger) multipoles $\ell$, nearby (distant) structures mainly contribute to $C_\ell$ because they appear larger (smaller) in the sky.
Especially for $\ell<10$, local structures at $z<0.01$ (corresponding to $r<30 \, h^{-1} \, \Mpc$) mainly determine the signal.

We comment that the analytical prediction of $C_\ell$ is less accurate for very small $\ell \, (<10)$, because it was derived under the flat-sky approximation. 
The accuracy of the Limber and flat-sky approximations  for projected galaxy clustering and weak-lensing statistics has been discussed in, e.g., \citet{Kilb2017} and \citet{Fang2020}.
Further studies are necessary to estimate the accuracy of these approximations in the DM statistics.

\subsection{Angular correlation function of the DM}

\begin{figure}
 \hspace{-1.2cm}
 \includegraphics[width=1.4 \columnwidth]{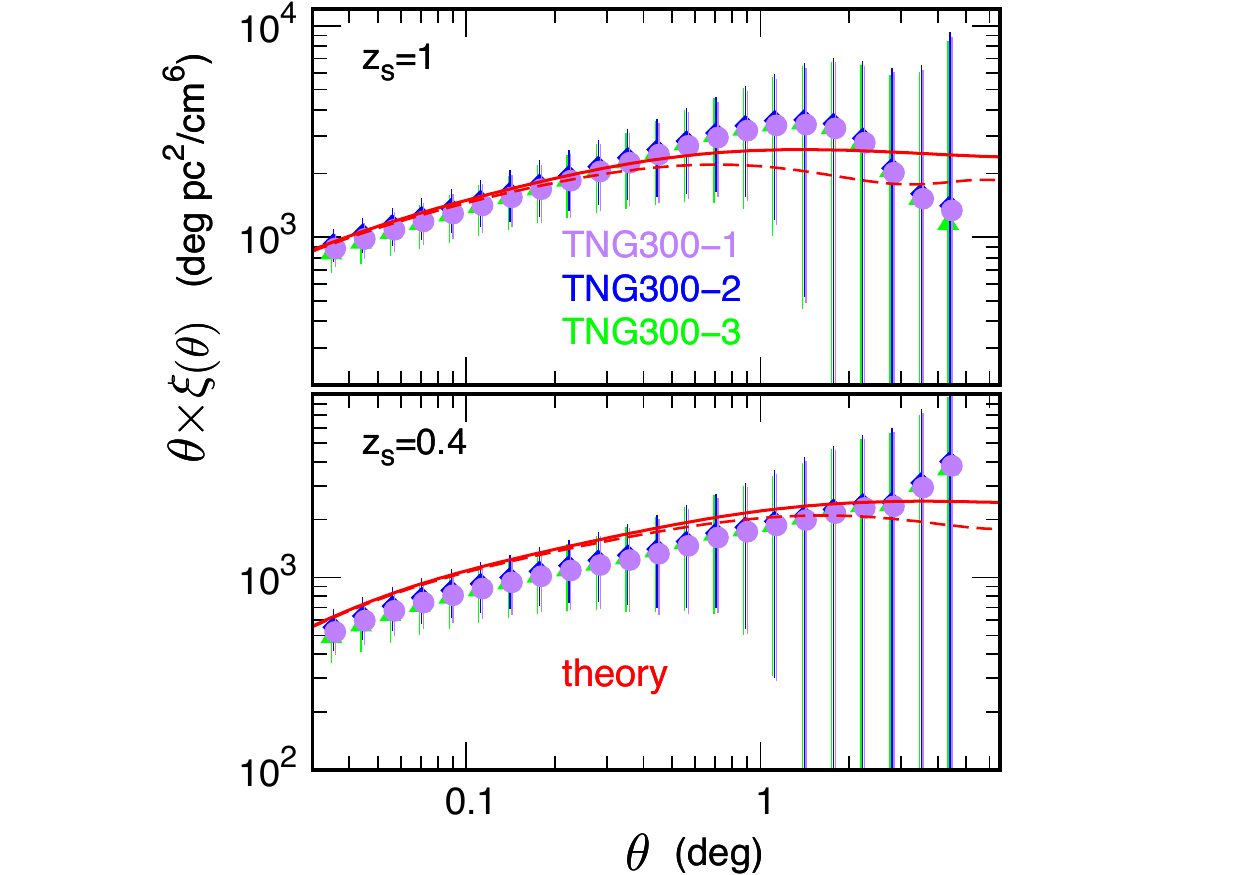}
 %\vspace{-0.5cm}
 \caption{Angular correlation functions of the DM at $z_{\rm s}=1$ (upper panel) and $0.4$ (lower panel). The purple, blue and green symbols are the TNG300-1, -2 and -3 results, respectively.  The mean and $1 \sigma$ error bars (= standard deviations) are measured from the $10$ mock maps. The field of view is $6 \times 6 \, {\rm deg}^2$, and the error bars scale as $[({\rm survey \, area})/(36 \, {\rm deg}^2)]^{-1/2}$.  The solid (dashed) red curves denote the analytical mean from Eq. (\ref{xi2}) (including the finite-simulation-box effect discussed in subsection 5.4).}
 \label{fig_xi}
\end{figure}

\begin{figure}
 \hspace{-1.2cm}
 \includegraphics[width=1.4 \columnwidth]{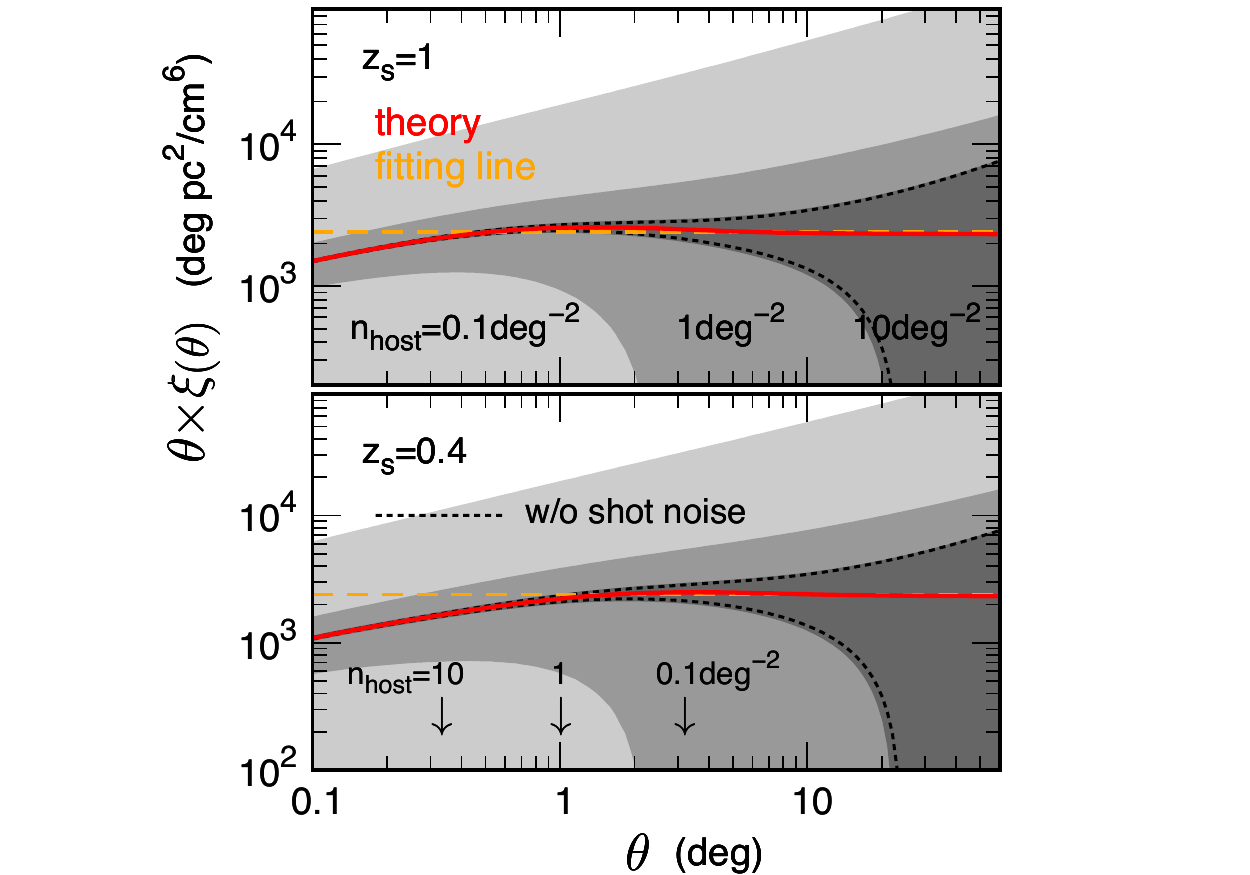}
 %\vspace{-0.5cm}
 \caption{Angular correlation functions of the DM for a full-sky measurement at $z_{\rm s}=1$ (upper panel) and $0.4$ (lower panel). The solid red curves denote the analytical mean (\ref{xi2}). The horizontal dashed-orange lines are the simple fit to the red curves given in Eq. (\ref{xi_fit}). The shaded area denotes the analytical standard deviation (\ref{xi_cov}) determined by the survey area and the shot noise from the host galaxies, with $\sigma_{\rm DM,host}=50 \, {\rm pc} \, {\rm cm}^{-3}$ for $n_{\rm host}=0.1 \, {\rm deg}^{-2}$ (light grey), $1 \, {\rm deg}^{-2}$ (grey), and $10 \, {\rm deg}^{-2}$ (dark grey). The dotted black curves denote the standard deviation without the shot noise (i.e., $n_{\rm host} \rightarrow \infty$). The standard deviation simply scales as $[({\rm survey \, area})/(4 \pi)]^{-1/2}$. The down arrows in the lower panel denote the average angular separation of the host galaxies (i.e., $n_{\rm host}^{-1/2}$) for $n_{\rm host}=0.1, 1$ and $10 \, {\rm deg}^{-2}$ from right to left.}
 \label{fig_xi_fullsky}
\end{figure}

%We also measured the angular correlation function of DM.
From Eq. (\ref{xi}), the correlation function in the i-th map is given by
\begin{align}
  \xi_{\rm i}(\theta;z_{\rm s}) = \frac{1}{N_{\rm pair}} \! \sum_{|\bftheta_1-\bftheta_2| \in \theta} & \! \left\{ \, {\rm DM}_{\rm i}(\bftheta_1;z_{\rm s}) - \overline{\rm DM}(z_{\rm s}) \right\} \nonumber \\
  & \times \left\{ \, {\rm DM}_{\rm i}(\bftheta_2;z_{\rm s}) - \overline{\rm DM}(z_{\rm s}) \right\},
  \label{xi_sim}
\end{align}
where the summation is done in the range $\theta-\Delta \theta/2<|\bftheta_1-\bftheta_2|<\theta+\Delta \theta/2$, and $N_{\rm pair}$ is the number of DM pairs in the bin-width $\Delta \log_{10} \theta=0.1$. 
The mean $\overline{\rm DM}$ is estimated from the $10$ maps.
Similarly to $C_\ell$, we measured $\xi_{\rm i}(\theta)$ for each of the $10$ maps to estimate its mean and variance among the maps.
We comment that $C_\ell$ and $\xi(\theta)$ are not independent but rather are related via the Fourier transform.

Figure \ref{fig_xi} plots the angular correlation functions at $z_{\rm s}=0.4$ and $1$.
The simulation results for the mean and standard deviations are obtained from the $10$ maps.
The standard deviations increase near the scale of the survey area ($=6$ deg) because the number of independent DM pairs decreases. 
The solid red curves are the analytical mean (\ref{xi2}). 
The dashed red curves include the effect of the finite simulation box, discussed in subsection 5.4.
%In calculating the integrate of Eq. (\ref{xi_cov}), we set the minimum multipole $\ell_{\rm min}=(2 \pi/6 \, {\rm deg}) \times \alpha = 60 \, \alpha$ where $\alpha$ is a factor of about unity.
The theory agrees well with the simulations
% at small scales ($\theta \lesssim 0.1 \, {\rm deg}$) 
but slightly overestimates them at $\theta \lesssim 1 \, {\rm deg}$ and $z_{\rm s}=0.4$.
This discrepancy %is also seen in $C_{\ell}$ at small $\ell$ (in Fig. \ref{fig_cl}), which 
may be caused by the sample variance. % or the inaccuracy of the flat-sky approximation.  
We comment that the simulation results between different values of $\theta$ are strongly correlated (see Eq. (\ref {xi_cov})). 
%(corresponding to the minimum wavenumber $k_{\rm min}=\ell_{\rm min}/r(z)$ in Eq. (\ref{xi2})).

Finally, Figure \ref{fig_xi_fullsky} plots the analytical correlation functions for a full-sky measurement. 
\rtrv{This figure shows the analytical results only but covers larger angular scales than Fig. \ref{fig_xi}.
The solid red curves are the theory (\ref{xi2}), which are the same as in Fig. \ref{fig_xi}.
Its asymptotic behavior at large $\theta$ can be described by a simple power law:}
\beq
 \xi(\theta;z_{\rm s}) \approx 2400  \left( \frac{\theta}{\rm deg} \right)^{-1} \, {\rm pc}^2 \, {\rm cm}^{-6}  ~~~{\rm for} ~\theta \gtrsim 1 \, {\rm deg}.
\label{xi_fit}
\eeq 
\rtrv{This is plotted by the horizontal dashed orange lines.}
%  at $\theta \gtrsim 1 \, {\rm deg}$ at the two source redshifts.
We checked that this fit works well at $z_{\rm s} \gtrsim 0.3$ (i.e., it is insensitive to $z_{\rm s}$, because such a large-scale signal is mainly determined by nearby structures, as shown in Fig. \ref{fig_cl_diffz}).
Note that Eq. (\ref{xi_fit}) simply scales as $\propto (f_{\rm e}/0.95)^2$ for an arbitrary $f_{\rm e}$.  
The shaded grey regions represent the standard deviation under the assumption of Gaussian density fluctuations (which is valid in the large-scale limit).
In this case, the covariance between $\xi(\theta_1;z_{\rm s})$ and $\xi(\theta_2;z_{\rm s})$ is given by \citep{Joachimi2008}
\begin{align}
 {\rm Cov}\left[ \xi(\theta_1;z_{\rm s}) \xi(\theta_2;z_{\rm s}) \right] &= \nonumber \\
  \frac{1}{S_{\rm W}} \int_0^\infty \! \frac{\ell d\ell}{\pi} & J_0(\ell \theta_1) J_0(\ell \theta_2) \left[  C_\ell(z_{\rm s}) + C_{\rm shot} \right]^2, 
  \label{xi_cov}
\end{align}
where $S_{\rm W}$ is the survey area in steradians.
The covariance is determined by the survey area at all scales and by the shot noise at small scales.
The diagonal element (i.e., $\theta_1=\theta_2$) corresponds to the variance. 
In this plot, the shot noise of the host galaxies is considered for the same three cases as in Fig. \ref{fig_cl}.
The dotted black curves denote the standard deviation without the shot noise. 
This figure suggests that the density $n_{\rm host}=10 \, {\rm deg}^{-2}$ is high enough to neglect the shot noise in the plotted range ($\theta > 0.1 \, {\rm deg}$).
For $n_{\rm host}=0.1 \, {\rm deg}^{-2}$, the shot noise affects the standard deviation even at very large scales ($\theta \gtrsim 10 \, {\rm deg}$).
%In this plot, the shot noise is not included.
At small scales, non-Gaussian fluctuations become important, and thus, the analytical prediction (\ref{xi_cov}) underestimates the results.
On larger scales, the standard deviation increases because there are fewer independent DM pairs in the full sky (i.e., the large-scale signal is limited by the cosmic variance).
The down arrows in the lower panel indicate the average angular separation for a given $n_{\rm host}$.  
Roughly, a signal larger than this scale can be measured. 
In the near future, when thousands of FRBs are available over the entire sky, we expect the correlation signal at $\theta \gtrsim 3 \, {\rm deg} \, (N_{\rm FRB}/4000)^{-1/2}$ to be detected (where $N_{\rm FRB}$ is the number of FRBs and the corresponding number density is $n_{\rm host} \simeq 0.1 \, {\rm deg}^{-2} (N_{\rm FRB}/4000)$).

%In this plot, we simply set $\ell_{\rm min}=0$.

Very recently, \cite{XB2020} reported the first detection of the angular two-point correlation of the DM using $112$ FRBs.
After subtracting the Milky Way contribution, they measured a statistical quantity---the so-called structure function $D(\theta)$---which is related to $\xi(\theta)$ as $D(\theta) = {\rm const}. -2 \xi(\theta)$.
Their result, plotted in their Fig. 3(b), is orders of magnitude larger than our analytical expectation (\ref{xi_fit}), although their error bars are still large.
More FRB samples are required to determine whether their signal is of cosmological origin or not. 
%These sample is too small to detect the signal from Fig. \ref{fig_xi_fullsky}.

\section{Discussion}

\subsection{Host-galaxy contribution}

So far we have not discussed the host-galaxy contribution, because there are two uncertainties in modelling it.
First, the host-galaxy properties show significant diversity among the $\sim 10$ currently identified host galaxies \citep[e.g.][]{Tendulkar2017,Procha2019,MacQ2020}.
For instance, the repeating source FRB 121102 is located in a dwarf galaxy, while four other sources identified by ASKAP are in massive galaxies \citep{Chatt2017,Bhan2020}.  
The spatial positions of the FRBs in their host galaxies also show variations from the centre to the outskirts.
Second, the resolution of TNG300 is not fine enough to resolve the inner structure of a host galaxy.
These large-box simulations are suitable for studying the cosmological distribution of free electrons, but to study the interiors of galaxies, finer-resolution (but smaller-box) runs---such as TNG50 and TNG100---are more suitable. 
\cite{Zhang2020} and \cite{Jaros2020} recently studied the host-galaxy contribution using TNG100.

When \cite{Pol2019} distributed the sources at a given $z_{\rm s}$ in their DM simulation, they compared two cases: (i) the sources are distributed randomly, and (ii) its distribution is proportional to the local density contrast.
They found that the latter significantly decreased the variance of the cosmological contribution, $\sigma_{\rm DM}^2$.  
Their results suggest that $\sigma_{\rm DM}$ also depends on host-galaxy properties such as its type (elliptical or spiral), mass or galaxy bias.  
%Here the host-galaxy distribution is specified by its bias factor, which mainly depends on the galaxy mass.
 %(or the bias factor  because that the host galaxy distribution is specified by the bias factor).  
More studies are needed on this topic, and we leave this for future work.

\subsection{Comparison with other hydrodynamic simulations}

Hydrodynamic simulations are the most reliable theoretical tool for studying the free-electron distribution. 
The cosmological DM has been studied using several simulations, such as Magneticum \citep{Dolag2015}, Illustris \citep{Jaros2019} and TNG300 (this work).
%In this paper, we measured the free-electron statistics using TNG300.
% but it is one example of hydrodynamic simulations.
Although these previous results are fairly consistent with ours (see section 5),  a more detailed quantitative comparison among various hydrodynamic simulations is desirable.
The free-electron distribution in halos depends strongly on the stellar and AGN feedback model that expels internal gas to the outside of a galaxy. 

\citet{Lim2020} recently studied the number-density profile of free electrons in halos with masses of $10^{12-14.5} \, M_\odot$ using three hydrodynamic simulations of Illustris, TNG300 and EAGLE \citep{Schaye2015}.
These simulations show a discrepancy of $\sim 30 \, \%$ at the halo radius $R_{\rm 500}$, as shown in their Figs. 5 and 6 (where $R_{500}$ is the radius within which the mean density is $500$ times larger than the mean cosmological background density).
The discrepancy is larger for a lower-mass halo, especially at smaller radius, because such halos are more sensitive to the feedback model.
For instance, Illustris predicts a low inner profile due to strong feedback.
In the halo model, $P_{\rm e}(k)$ at small scales ($k \gtrsim 1 \, h \, \Mpc^{-1}$) is determined by the halo mass function and the free-electron density profile in the halos.
%The one-halo term is proportional to the square of the Fourier transform of the profile.
Therefore, a similar level of discrepancy is probably present in $P_{\rm e}(k)$.

%Because the feedback affects the small-scale $P_{\rm e}(k)$, the DM statistics may constrain the feedback models.
Because the DM variance is sensitive to $k^2 P_{\rm e}(k)$ around the peak ($k \approx 1$--$10 \, h \, \Mpc^{-1}$), the uncertainty in the feedback model may affect the variance.
%the variance may constrain the feed back models
The angular power spectrum of the DM at small scales ($\ell > 10^3$) is also sensitive to the feedback.
However, because its small-scale signal is strongly contaminated by the shot noise (see Fig. \ref{fig_cl}), the feedback effect will be difficult to observe in $C_\ell$. %in the near future.}
%but a high number density of FRBs ($n_{\rm host}>10 \, {\rm deg}^{-2}$) is required to detect it over the shot noise level (see
%but and it is probably difficult to detect it.}

\section{Conclusions}

We have investigated the basic statistics of the cosmological dispersion measure (DM) using the state-of-the-art hydrodynamic simulations, IllustrisTNG300.
%As DM is the projected free-electron number density, 
Our main purpose is to provide an analytical model for data analysis on the DM statistics.
 
First, we measured the free-electron fraction $f_{\rm e}(z)$ and its power spectrum $P_{\rm e}(k;z)$ from TNG300, which are 
%The free-electron abundance and its spatial distribution are 
key ingredients in the DM statistics.
It turns out that $P_{\rm e}(k;z)$ is consistent with the dark-matter-only power spectrum $P_{\rm dmo}(k;z)$ at large scales ($k \lesssim 1 \, h \, \Mpc^{-1}$), but it is strongly suppressed at small scales ($k \gtrsim 1 \, h \, \Mpc^{-1}$) owing to stellar and AGN feedback.
As a result, the free-electron fluctuations on scales $\approx 1 \, \Mpc$ contribute most to the DM variance (because $k^2 P_{\rm e}(k;z)$ has a peak around that scale).
To model $P_{\rm e}(k;z)$, we introduced the free-electron bias factor defined by $b^2_{\rm e}(k;z)=P_{\rm e}(k;z)/P_{\rm dmo}(k;z)$.
We then provided simple fitting functions calibrated over a wide range of scales and epochs: $f_{\rm e}(z)$ for $z=0$--$8$ in Eq. (\ref{fe_fit}) and $b_{\rm e}(k;z)$ for $k<10 \, h \, \Mpc^{-1}$ and $z=0$--$5$ in Eq. (\ref{bias_fit}).
These fitting functions will be useful for future statistical analyses of the free-electron distribution.

Next, we prepared $10$ mock sky maps of the DM using the TNG300 data, based on standard ray-tracing techniques.
We then measured various DM statistics, such as its mean and variance, PDF of the DM, PDF of the source redshift $z_{\rm s}$ for a given DM, angular power spectrum and angular correlation function. 
We calculated the analytical predictions using the fitting formulas for $f_{\rm e}(z)$ and $P_{\rm e}(k;z)$ and then validated them against the mock DM measurements. 
Basic statistics such as the mean, variance and PDF of the DM were consistent with previous work.
The PDF of the DM is highly skewed, while the PDF of the $z_{\rm s}$ is well approximated by a Gaussian.
We provided a source redshift--DM relation---$z_{\rm s}=z_{\rm s}({\rm DM})$ in Eq. (\ref{z-DM_fit})---which helps in identifying the host galaxies of FRBs from the measured DMs. 
The angular correlation function was also computed in subsection 5.5, and we expect it to be detected when thousands of FRBs are available in the coming years. 
  
Throughout this paper,  we compared the TNG300 results with three resolution runs to see the numerical convergence.
We confirmed that our conclusions do not depend on the resolution, because all the runs resolve the dominant length scale of the free-electron fluctuations ($\approx 1 \, \Mpc$) sufficiently.  % which mainly affects the DM variance).   
Even so, because the gas distribution in halos is sensitive to the feedback model, quantitative comparisons with other hydrodynamic simulations are required for further systematic checks.
%and confirmed that our results are converged.
%\rtrv{Although our model ingredients (the ionised fraction $f_{\rm e}$ and the power spectrum $P_{\rm e}$) were calibrated using TNG300, 
The presented analytical model for the DM statistics will be updated easily by re-calibrating the fitting functions for $f_{\rm e}(z)$ and $P_{\rm e}(k;z)$ using more accurate future hydrodynamic simulations.

\section*{Acknowledgements}

We thank the IllustrisTNG team very much for making their simulation data publicly available.
We thank Atsushi J. Nishizawa and Shinpei Nishioka for their useful comments. 
This work is supported by MEXT/JSPS KAKENHI Grant Numbers 20H04723, 17H01131 (RT), 20H01901, 20H01904, 20H00158, 18H01213, 18H01215, 17H06357, 17H06362, 17H06131 (KI).

\section*{Data Availability}

The TNG300 simulation data is available at \url{https://www.tng-project.org}.
The measurement data underlying this article will be shared on reasonable request to the first author.

%%%%%%%%%%%%%%%%%%%%%%%%%%%%%%%%%%%%%%%%%%%%%%%%%%



\bibliographystyle{mnras}
\bibliography{refs} % if your bibtex file is called example.bib

%%%%%%%%%%%%%%%%%%%%%%%%%%%%%%%%%%%%%%%%%%%%%%%%%%

%%%%%%%%%%%%%%%%% APPENDICES %%%%%%%%%%%%%%%%%%%%%

\appendix

%\section{Derivation of the two-point statistics}

%This appendix presents derivation of the two-point statics of DM.
%The correlation function of DM is rewritten from Eqs.(\ref{DM_fluct}) and (\ref{xi}) as 
%\beq
%  \xi(\theta_{12};z_{\rm s}) =  \int_0^{z_{\rm s}} \!\! \frac{c dz}{H(z)} W(z)  \int_0^{z_{\rm s}} \!\! \frac{c dz^\prime}{H(z^\prime)} W(z^\prime)  \langle \delta_{\rm e}(\bfr;z)   \delta_{\rm e}(\bfr^\prime;z^\prime) \rangle  , 
%\eeq

%%%%%%%%%%%%%%%%%%%%%%%%%%%%%%%%%%%%%%%%%%%%%%%%%%

% Don't change these lines
\bsp	% typesetting comment
\label{lastpage}
\end{document}